\documentclass[aps,pre,reprint,superscriptaddress]{revtex4-1}

\pdfoutput=1

\usepackage{graphicx}
\usepackage{amssymb}
\usepackage{amsmath}
\usepackage{xcolor}
\usepackage{pdfpages}
\usepackage{pgffor}
\usepackage{multirow}
\usepackage{gensymb}

\allowdisplaybreaks
\usepackage{placeins} 

\makeatletter
\AtBeginDocument{\let\LS@rot\@undefined}
\makeatother

\begin{document}

\title{Effect of Particle Shape on Stratification in Drying Films of Binary Colloidal Mixtures}


\author{Binghan Liu}
\affiliation{Department of Physics, Center for Soft Matter and Biological Physics, and Macromolecules Innovation Institute, Virginia Tech, Blacksburg, Virginia 24061, USA}
\author{Gary S. Grest}
\affiliation{Sandia National Laboratories, Albuquerque, NM 87185, USA}
\author{Shengfeng Cheng}
\email{chengsf@vt.edu}
\affiliation{Department of Physics, Center for Soft Matter and Biological Physics, and Macromolecules Innovation Institute, Virginia Tech, Blacksburg, Virginia 24061, USA}
\affiliation{Department of Mechanical Engineering, Virginia Tech, Blacksburg, Virginia 24061, USA}

\date{\today}

\begin{abstract}
The role of particle shape in evaporation-induced auto-stratification in dispersed colloidal suspensions is explored with molecular dynamics simulations of mixtures of solid spheres, aspherical particles (disks, rods, tetrahedra, and cubes), and hollow spheres. A unified framework is proposed for the stratification phenomena in systems that feature size or shape dispersity on the basis of two processes: diffusion and diffusiophoresis. In general, diffusion favors the accumulation of particles that diffuse more slowly (e.g., particles of larger sizes) at the evaporation front. However, particles with larger surface areas (e.g., particles of larger sizes or lower sphericity) have larger diffusiophoretic mobilities and are more likely to be driven away from the evaporation front by the concentration gradients of other particles with smaller surface areas. In the case of a bidisperse colloidal suspension containing small and large solid spheres studied in most of the work reported in the literature, the competition between the two leads to the so-called ``small-on-top'' stratification when the suspension is rapidly dried, as diffusiophoresis dominates near the interface. In this work, we employ a computational model of composite particles that mimics the Rouse model of polymers, where the diffusion coefficient of a particle is inversely proportional to its mass. For a mixture of solid spheres and aspherical particles or hollow spheres with similar masses, the diffusion contrast is reduced and the solid spheres are always enriched at the evaporation front, as they have the smallest surface area for a given mass and therefore the lowest diffusiophoretic mobility. The unified framework is further corroborated with a case study of a mixture of solid and hollow spheres having the same outer radius and thus the same surface area. Because in this case the diffusiophoretic contrast is suppressed, this system is dominated by diffusion. As a result, the solid spheres, which have a larger mass and thus a smaller diffusion coefficient, are found to accumulate at the evaporation front. Finally, in a mixture of thin disks and long rods that differ significantly in shape but have similar masses and surface areas, both diffusion and diffusiophoresis contrasts are suppressed and the mixture does not stratify at all.
\end{abstract}

\maketitle

\section{Introduction}\label{intro}

Drying a colloidal suspension is a widely used technique for the manufacture of ordered structures such as photonic crystals,\cite{Cai2021CSR} the production of nanocomposite materials,\cite{Kumar2017JCP} the coating of a surface,\cite{KeddieRouth2010} and the production of a thin film.\cite{HeJinbo2010} As the solvent evaporates from the suspension, the complex interplay between diffusion, evaporation-induced particle advection and convection, interactions among particles, substrate, liquid-vapor interface and potentially external fields, and various gradients (e.g., in temperature and concentration) emerging during evaporation dictates the particle transport and distribution in the drying suspension and the structure they eventually assemble into.\cite{Brinker2004} Drying suspensions therefore provide a versatile platform for exploring non-equilibrium soft matter physics \cite{Routh2013} and a deeper understanding of the associated diffusive, phoretic, and self-assembly processes will lead to meticulous control of drying conditions to yield desired materials and structures.\cite{Brinker1999, Russel2011, Zhang2023JEECS}

In recent years, self-stratification phenomena in rapidly dried colloidal suspensions possessing sufficient size dispersity have attracted great attention, as it can potentially lead to a one-pot route for creating multilayered coatings and film materials quickly, efficiently, and cost-effectively.\cite{Zhou2017AdvMater, Schulz2018, LiYifan2022, He2023ChemMater} In particular, bidispere suspensions containing large and small spherical colloids have been studied extensively and various outcomes have been achieved, including large-on-top stratified, small-on-top stratified, sandwiched, and uniform distributions of particles.\cite{Cardinal2010, Nikiforow2010, Trueman2012JCIS, Trueman2012Langmuir, Atmuri2012, Fortini2016, Martin-Fabiani2016, Makepeace2017, Schulz2021, Palmer2023, Carr2018, LiuXiao2018, LiuWeiping2019, Carr2024, Dong2020SM, Sofroniou2024, Cusola2018, Romermann2019, Coureur2023, Hooiveld2023, Hooiveld2025, LiSiyu2023, Tiwari2023} Here the term ``on-top'' indicates that the corresponding particles gather at the evaporation front. In the most simplified model (e.g., hard spheres dispersed in a thin film suspension undergoing one-dimensional evaporation along the normal direction of the film), the dominant factor affecting particle distribution is the competition between their advection imposed by the receding liquid-vapor meniscus and the diffusion of the particles.\cite{Fortini2016, Zhou2017, Sear2017, Sear2018, Howard2017, Statt2017, Tang2018Langmuir, Schulz2021, Rees-Zimmerman2021, Rees-Zimmerman2024} At fast solvent evaporation, advection can cause the particles to accumulate at the evaporation front and develop concentration gradients, which then induce diffusiophoretic motion of the particles. Therefore, the competition between diffusion and diffusiophoresis dictates the outcome of film drying. In general, diffusion promotes the accumulation of particles that diffuse more slowly at the top, which in the case of bidisperse spheres are the larger particles, while diffusiophoresis favors the enrichment of the smaller particles at the evaporation front,\cite{Rees-Zimmerman2021, Rees-Zimmerman2024} because the larger particles have larger diffusiophoretic mobilities and are more likely to be pushed away from the evaporation front. This is the reason why both the evaporation rate and the initial volume fraction of the smaller particles need to be sufficiently high to produce the so-called ``small-on-top'' stratification as both are necessary for the smaller particles to develop concentration gradients so that diffusiophoresis can take effect.\cite{Zhou2017, Sear2017, Sear2018, Tang2018Langmuir}

Despite that the simplest model captures the main physics of film drying, realistic systems are always more complicated with many factors at play.\cite{Routh2013, Zahedi2018} Various attempts have been made to understand the role of interparticle interactions beyond exclude volume repulsion,\cite{Atmuri2012, XiaoMing2019, Samanta2020, Tinkler2022, Tatsumi2023, Sofroniou2024, Hooiveld2025} hydrodynamic interactions,\cite{Statt2017, Howard2020} the presence of additives (e.g., salts, surfactants, and polymers),\cite{Lee2020SM, Tinkler2022, Rees-Zimmerman2024, Hooiveld2025} the dispersity and size ratio of particles,\cite{Martin-Fabiani2016, Trueman2012JCIS, Fortini2017, LiuWeiping2019, XiaoMing2019, LiuWeiping2019, Carr2024, Hooiveld2025} the interfacial activity of particles,\cite{Kim2024Small} the wetting property of particles,\cite{Li2020MaterHoriz, He2023ChemMater} the surface chemistry of particles,\cite{Dong2020SM, Samanta2020} and the solvent viscosity.\cite{Murdoch2023} However, one important aspect of particle properties, their shape, has not been systematically explored for its role in evaporation-induced auto-stratification. The purpose of the present study is to fill this gap.

Onsager was the first to note that shape has an effect on interactions between colloidal particles.\cite{Onsager1949} More recently, the effects of particle shape on the drying behavior of soft matter solutions have also been documented in the literature. Sau and Murphy showed that particle shape strongly influenced the self-assembly patterns in aqueous solutions of cetyltrimethylammonium bromide-coated gold nanocrystals upon solvent evaporation, as the shape dictates the direction-specific interactions between the coated particles.\cite{Sau2005} Koike \textit{et al.} simulated the drying process of colloidal rod suspensions on the basis of the Langevin equation and found that the final structure depends on the aspect ratio of the rods.\cite{Koike2008} Hodges \textit{et al.} found structural differences in the dried coffee-ring deposits left by drops of spherical Ludox silica or disk-shaped laponite clay nanoparticles after solvent evaporation.\cite{Hodges2010} Yunker \textit{et al.} showed that the coffee-ring effect can be suppressed with ellipsoidal particles, as their aspherical shape leads to long-range capillary attractions among the particles at the water-air interface.\cite{Yunker2011Nature} Yunker \textit{et al.} also found that particle shape influenced the structure of final deposition and the bending rigidity of colloidal monolayer membranes produced by drying aqueous drops confined between glass plates.\cite{Yunker2012PRL} In another work, Yunker \textit{et al.} showed that the growth dynamics at the edges of evaporating drops falls into different universality classes depending on the anisotropy of the suspended particles.\cite{Yunker2013PRL}

Th\'{e}rien-Aubin \textit{et al.} found that drying mixed suspensions of cellulose nanorods and latex nanospheres resulted in films with chiral nematic layers enriched with cellulose nanorods alternating with planar disordered layers of latex nanospheres.\cite{Therien-Aubin2015} Dugyala \textit{et al.} found that crack patterns after drying change from being along the radial direction to being circular when the particles suspended in sessile drops are switched from spherical to ellipsoidal.\cite{Dugyala2016} Gong \textit{et al.} combined experiment and Monte Carlo simulations to investigate the self-assembly process in suspensions of nanocrystals with various shapes induced by sedimentation and solvent evaporation, and found that the ultimate coherence length of the resulting superlattices strongly depends on the nanocrystal shape.\cite{Gong2017} Park \textit{et al.} used multi-speckle diffusing wave spectroscopy to probe the Brownian motion of spherical and ellipsoidal polystyrene particles in evaporating drops and found that ellipsoidal particles move more slowly and are less readily to form coffee-ring deposits.\cite{Park2018CPS} Gracia-Medrano-Bravo \textit{et al.} recently studied the drying behavior of solutions containing both colloidal particles and polymers and showed that compared with spherical particles, plate-like particles lead to an expansion of the diffusive regime, where the particles are uniformly distributed in the polymer film after drying.\cite{Gracia-Medrano-Bravo2020} They later found numerically an increase in the evaporation regime, where the particles accumulate at the top of the polymer film, when the aspect ratio of the aspherical particles is reduced.\cite{Gracia-Medrano-Bravo2021}

Recently, Sui showed that platelet–sphere mixtures stratify during sedimentation by numerically solving the sedimentation–diffusion equations.\cite{Sui2019SM} The goal of the present study is to reveal the role of particle shape in evaporation-induced stratification by investigating colloidal suspensions that contain mixtures of solid spheres and aspherical particles or hollow spheres. Their drying processes are modeled with molecular dynamics simulations based on the moving interface method, where the solvent is treated as an implicit viscous background and the liquid-vapor interface is treated as a (movable) confining potential for the suspended particles.\cite{Tang2022} Composite particles with various shapes are constructed with Lennard-Jones beads.\cite{Nguyen2019} Each bead is coupled to the viscous background via a Langevin thermostat, whose strength thus determines the diffusion coefficients of the particles. In this method, hydrodynamic interactions between particles are neglected, as previous work showed that for colloidal suspensions, such an implicit solvent model can be reasonably matched to an explicit solvent model.\cite{Tang2019JCP_compare} However, it should be noted that a similar match might be questionable for polymer solutions, where hydrodynamic interactions seem to play a much more important role.\cite{Statt2018, Howard2020} More studies are needed to fully elucidate the effects of hydrodynamic interactions on the drying outcomes of polymer solutions, colloidal suspensions, and their mixtures.\cite{Kundu2022}

The remainder of this paper is organized as follows. The simulation methodology is described in Sec.~\ref{sec:md}. The simulation results are discussed in Sec.~\ref{sec:res}. Finally, the findings are summarized and concluded in Sec.~\ref{sec:conc}.

\section{Simulation Methodology}\label{sec:md}

\begin{figure}[htbp] 
    \centering
    \includegraphics[width=0.35\textwidth]{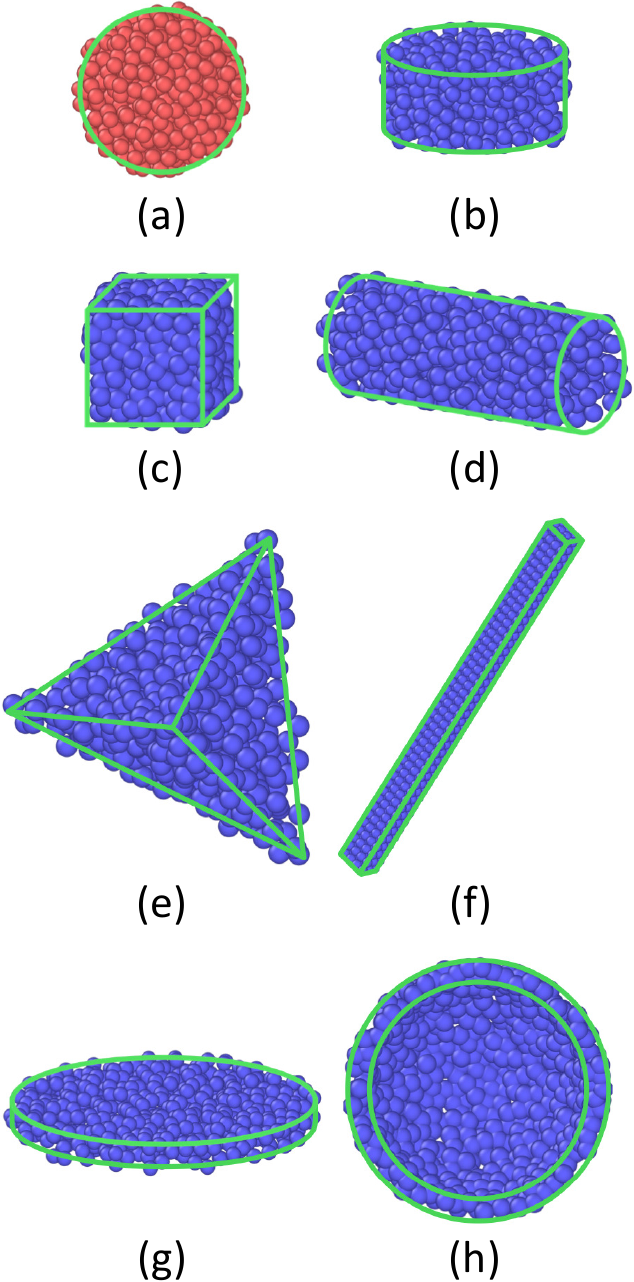}
    \caption{Composite particles with various shapes: (a) solid sphere, (b) thick disk, (c) cube, (d) short rod, (e) regular tetrahedron, (f) long rod, (g) thin disk, (h) hollow sphere. Except for the hollow sphere, the other particles are ordered with decreasing sphericity. Particles in (b)-(h) correspond to the type-2 particles in systems S1-S7 in Table~\ref{table:systems}.}
    \label{fig:comp_particle_shape}
\end{figure}

A composite rigid-body model is used to simulate colloidal particles with various shapes,\cite{Nguyen2019} on the basis of point beads interacting with each other via the standard Lennard-Jones (LJ) 12-6 potential,
\begin{equation}\label{eqn:lj_pot}
    U_\text{LJ}(r_{ij})=4\epsilon\left[\left(\frac{\sigma}{r_{ij}}\right)^{12} - \left(\frac{\sigma}{r_{ij}}\right)^{6}\right]~,
\end{equation}
where $\epsilon$ and $\sigma$ set the energy and length scale of the interaction, respectively, and $r_{ij}$ is the distance between two beads labeled $i$ and $j$. The LJ potential is truncated and shifted to $0$ at $r_c = 2^{1/6} \sigma$, making the interactions between colloidal particles fully repulsive to prevent agglomeration. Setting the mass of an LJ bead as $m$, the LJ unit system can be defined in terms of $m$, $\epsilon$, and $\sigma$. For example, the LJ unit of time is $\tau \equiv \sqrt{m \sigma^2/\epsilon}$.

To construct composite particles with different shapes, an equilibrium LJ melt is prepared with a number density of $1.0\sigma^{-3}$ at temperature $1.0\epsilon/k_\text{B}$, where $k_\text{B}$ is the Boltzmann constant. Spherical and asphrerical composite particles are then carved out of this melt, including solid spheres, hollow spheres, regular tetrahedra, cubes, rods, and disks. The solid sphere has a radius of $5\sigma$ and contains $527$ LJ beads. The hollow sphere, which has an outer radius of $7\sigma$ and an inner radius of $6\sigma$, consists of $537$ LJ beads. The dimensions of aspherical particles (e.g., regular tetrahedra and cubes) are chosen in such a way that they contain approximately the same number ($\sim 527\pm 10$) of LJ beads as in the solid sphere. The diameter and length of the disks or rods are chosen to produce various aspect ratios, defined as the ratio of the length of the cylinder to its cross-sectional diameter, while keeping the mass close to that of the solid sphere. Two types of disk are prepared with an aspect ratio of 0.084 (thin disk) and 0.5 (thick disk), respectively. A short and thick rod with an aspect ratio of 4 is also carved from the melt. To explore the role of rod length, another long and thin rod with a much larger aspect ratio of 20 is prepared. However, it is difficult to maintain the cylindrical shape by carving a long rod out of an LJ melt. As a result, the long and thin rod is carved from a simple cubic lattice of LJ beads at a number density of $1.0\sigma^{-3}$. The images of the composite particles studied in this paper are shown in Fig.~\ref{fig:comp_particle_shape} and their properties are summarized in Table~\ref{table:systems}.

One measure of the shape of an object is sphericity, which is defined as
\begin{equation}
    \Psi = \frac{\pi^{1/3}\left( 6V_p\right)^{2/3}}{A_p}~,
\end{equation}
where $V_p$ is the volume of the object and $A_p$ its surface area. For a sphere, $\Psi = 1$, while for aspherical objects, $\Psi < 1$. All of the (solid) composite particles described previously have approximately the same volume. Therefore, their sphericity is roughly inversely proportional to their surface area. Here, to quantify $\Psi$, we use the volume of the solid sphere with a radius of $5\sigma$ as $V_p$ and, as a result, the value of $\Psi$ reported in Table~\ref{table:systems} is inversely proportional to the area ($A_2$) of the aspherical composite particle (i.e., the type-2 particle in Table~\ref{table:systems}).



\begin{table*}[htbp] 
\caption{All mixtures (of type 1 and type 2 particles) examined in this paper. The rest columns are for the following physical quantities: mass ($M$), surface area ($A$), sphericity ($\Psi$), diffusion coefficient ($D$), and P\'{e}clet number (Pe) under the fast-drying condition ($v_{e} = 1.311\times 10^{-2}\sigma/\tau$). Subscripts 1 and 2 are for type 1 and type 2 particles, respectively.}
\begin{tabular}{|c|c|c|c|c|c|c|c|c|c|c|c|c|}
    \hline
    System & ~~~Type 1~~~ & ~~~Type 2~~~ & $M_{1}$ ($m$) & $M_{2}$ ($m$) & $A_{1}$ ($\sigma^{2}$) & $A_{2}$ ($\sigma^{2}$) & $A_{2}/A_{1}$ & $\Psi_{2}$ & $D_{1}$ ($\sigma^{2}\tau^{-1}$) & $D_{2}$ ($\sigma^{2}\tau^{-1}$) & ~~~Pe$_{1}$~~~ & ~~~Pe$_{2}$~~~ \\ \hline
    
    $ \text{S1} $ & {\raisebox{-0.15\height}{\includegraphics[width=0.35cm]{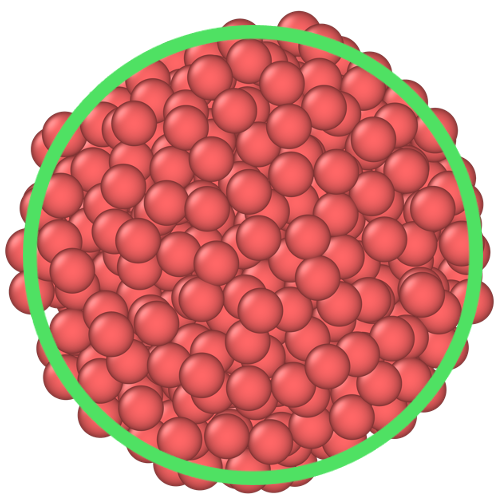}}} & {\raisebox{-0.2\height}{\includegraphics[width=0.5cm]{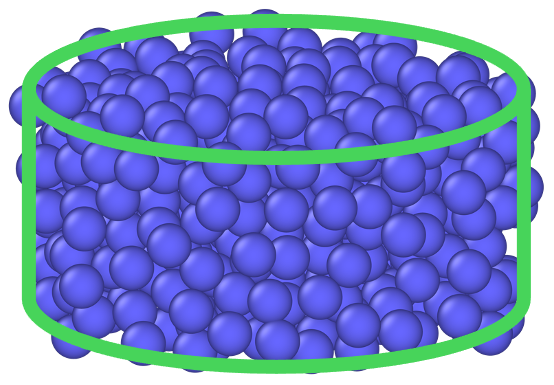}}} & $~527$ & $~522$ & $314.16$ & $~380.13$ & 1.21 & 0.825 & $0.0256$ & $0.0283$ & $408.1$ & $382.8$  \\ \hline
    
    $ \text{S2} $ & {\raisebox{-0.15\height}{\includegraphics[width=0.35cm]{sp.png}}} & {\raisebox{-0.2\height}{\includegraphics[width=0.4cm]{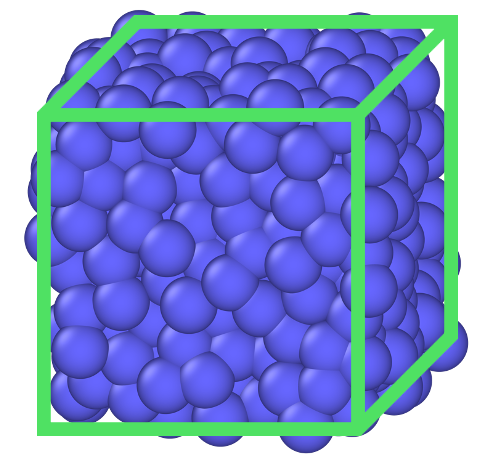}}} & $~527$ & $~527$ &  $314.16$ & $~384.00$ & 1.22 & 0.822 & $0.0264$ & $0.0263$ & $397.3$ & $398.8$ \\ \hline
    
    $ \text{S3} $ & {\raisebox{-0.15\height}{\includegraphics[width=0.35cm]{sp.png}}} & {\raisebox{-0.2\height}{\includegraphics[width=0.6cm]{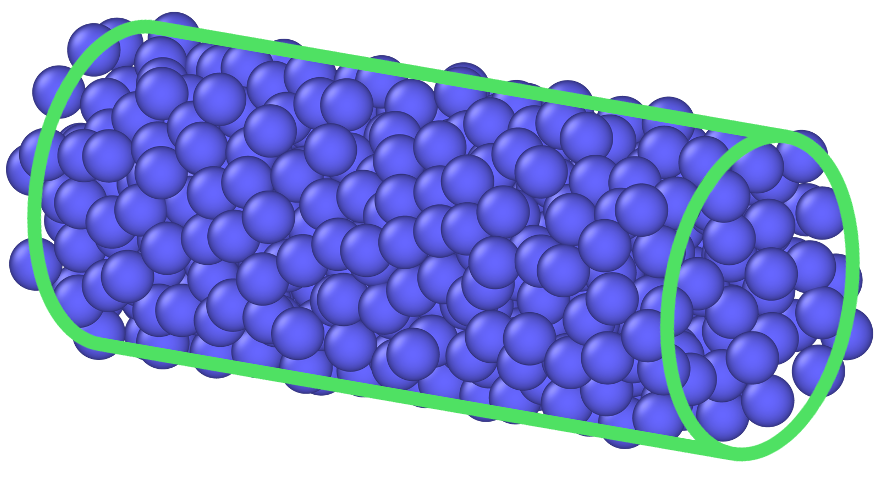}}} & $~527$ & $~533$ &  $314.16$ & $~433.27$ & 1.38 & 0.734 & $0.0279$ & $0.0291$ & $409.7$ & $370.6$ \\ \hline
    
    $ \text{S4} $ & {\raisebox{-0.15\height}{\includegraphics[width=0.35cm]{sp.png}}} & {\raisebox{-0.3\height}{\includegraphics[width=0.5cm]{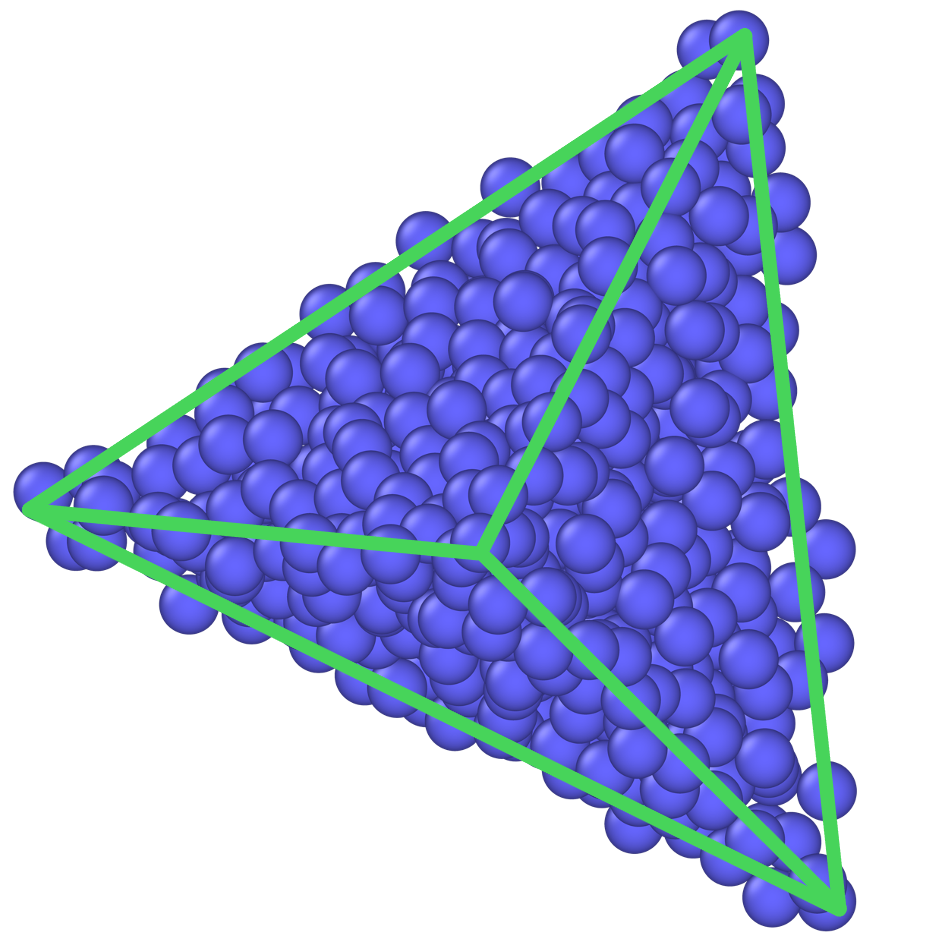}}} & $~527$ & $~518$ &  $314.16$ & $~468.20$ & 1.49 & 0.667 & $0.0275$ & $0.0278$ & $381.4$ & $377.3$ \\ \hline
    
    $ \text{S5} $ & {\raisebox{-0.15\height}{\includegraphics[width=0.35cm]{sp.png}}} & {\raisebox{0.02\height}{\includegraphics[width=1.0cm]{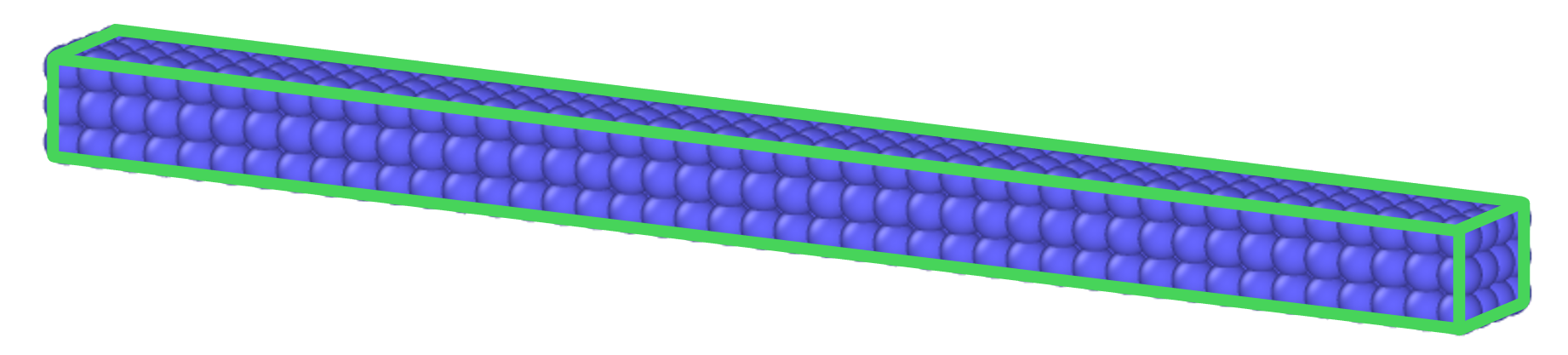}}} & $~527$ & $~516$ &  $314.16$ & $~660.56$ & 2.10 & 0.471 & $0.0251$ & $0.0285$ & $417.9$ & $368.0$ \\ \hline
    
    $ \text{S6} $ & {\raisebox{-0.15\height}{\includegraphics[width=0.35cm]{sp.png}}} & {\raisebox{-0.08\height}{\includegraphics[width=0.65cm]{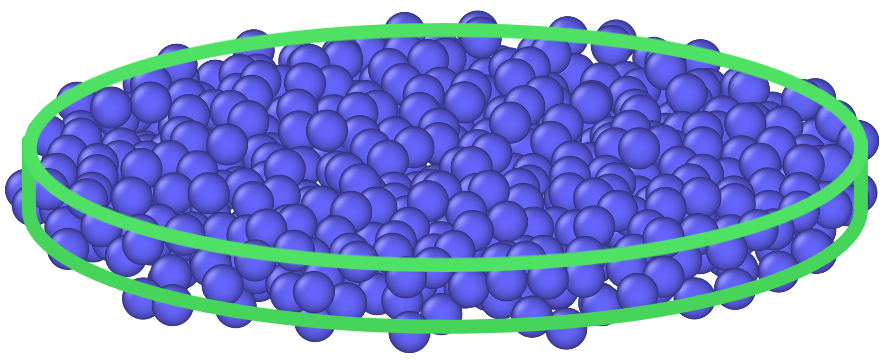}}} & $~527$ & $~520$ & $314.16$ & $~726.56$ & 2.31 & 0.430 & $0.0257$ & $0.0274$ & $375.9$ & $360.4$ \\ \hline
    
    $ \text{S7} $ & {\raisebox{-0.15\height}{\includegraphics[width=0.35cm]{sp.png}}} & {\raisebox{-0.26\height}{\includegraphics[width=0.45cm]{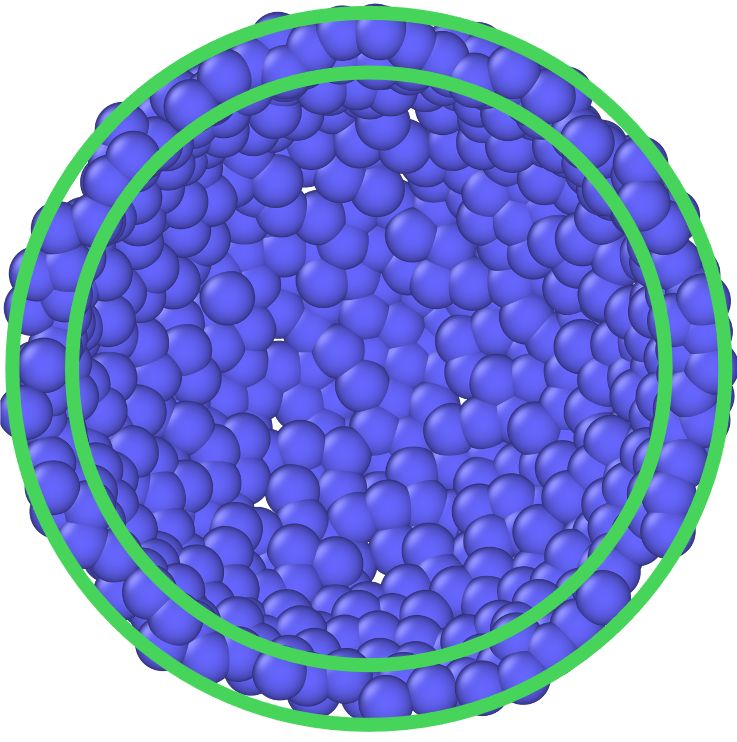}}} & $~527$ & $~537$ &  $314.16$ & $~615.75$ & 1.96 & $-$ & $0.0264$ & $0.0215$ & $397.3$ & $487.9$ \\ \hline
    
    $ \text{S8} $ & {\raisebox{-0.27\height}{\includegraphics[width=0.45cm]{sp.png}}} & {\raisebox{-0.3\height}{\includegraphics[width=0.55cm]{hollow.png}}} & $1147$ & $1134$ & $530.93$ & $1256.64$ & 2.37 & $-$ & $0.0159$ & $0.0139$ & $659.6$ & $754.5$\\ \hline

    $ \text{S9} $ & {\raisebox{-0.15\height}{\includegraphics[width=0.35cm]{sp.png}}} & {\raisebox{-0.15\height}{\includegraphics[width=0.35cm]{hollow.png}}} & $~527$ & $~261$ & $314.16$ & $~314.16$ & 1.00 & $-$ & $0.0244$ & $0.0485$ & $430.5$ & $216.2$\\ \hline
    
\end{tabular}
\label{table:systems}
\end{table*}

Seven mixtures are first prepared (systems S1-S7 in Table~\ref{table:systems}), each consisting of the solid spheres of radius $5.0\sigma$ and another type of composite particles. Each mixture is placed in a rectangular simulation box of dimensions $200\sigma \times 200\sigma \times 800\sigma$ in the $x$, $y$ and $z$ directions, respectively. In the $xy$ plane, periodic boundary conditions are utilized, while along the $z$ axis, the particles are confined between two walls, one located at $z=0$ and the other at $z=800\sigma$. The interaction between an LJ bead and each wall is given by an LJ 9-3 potential,
\begin{equation} \label{eqn:wall}
    U_\text{w}(r_\perp)=\epsilon \left[\frac{2}{15}\left(\frac{\sigma}{r_\perp}\right)^{9}
    -\left(\frac{\sigma}{r_\perp}\right)^{3}\right]~,
\end{equation}
where $r_\perp$ is the distance between the LJ bead and the wall. In each mixture, the number of solid spheres is the same as the number of aspherical particles (i.e., tetrahedra, cubes, disks, or rods); both have the same initial volume fraction of $0.02$. The initial volume fraction of the hollow spheres in system S7 is $0.043$.

For purposes explained later, two additional mixtures of solid and hollow spheres, namely systems S8 and S9, are also prepared. In system S8, solid spheres have a radius of $6.5\sigma$ and contain $1147$ LJ beads; hollow spheres have an outer radius of $10\sigma$ and an inner radius of $9\sigma$ and contain $1134$ LJ beads. The initial volume fraction in system S8 is $0.012$ for the large solid spheres and $0.038$ for the large hollow spheres. In system S9, solid spheres have a radius of $5.0\sigma$; hollow spheres have an outer radius of $5.0\sigma$ and an inner radius of $4.0\sigma$ and contain $261$ LJ beads. The solid and hollow spheres in S9 have the same surface area but different mass. All nine systems are summarized in Table~\ref{table:systems}. The aspherical particles are ordered by their sphericity and the corresponding mixtures with the solid spheres (radius $5.0\sigma$) are labeled S1 to S6, respectively, while the three systems that involve hollow spheres are labeled S7 to S9.

All molecular dynamics (MD) simulations reported here are performed with the Large-scale Atomic/Molecular Massively Parallel Simulator (LAMMPS).\cite{LAMMPS} The Newtonian equations of motion are integrated in a time step of $0.01\tau$ using the Velocity Verlet algorithm. A Langevin thermostat is applied to each LJ bead to keep the system at a constant temperature of $1.0\epsilon/k_\text{B}$. The damping time of the thermostat is tuned to control the diffusion coefficient of the composite particles. The current study is therefore based on an implicit solvent model, where the solvent is treated as a viscous background via the Langevin thermostat. For all simulations reported here, the damping time is set at $\Gamma=15.7\tau$.\cite{Tang2019JCP_compare}

For each mixture, the particles are initially placed on a pre-defined lattice to avoid heavy overlaps. In subsequent MD runs, each composite particle is held as a rigid body. A soft potential is used to separate the composite particles with any overlapped LJ beads. Then the soft potential is turned off and the LJ potential is turned on, followed by an equilibration run to produce a uniform particle suspension.

The evaporation process of each mixture solution is modeled with the moving interface method.\cite{Tang2022} In this method, the role of a liquid-vapor interface is replaced by a potential barrier that confines all the solute particles in the suspension phase. The repulsive half of a harmonic potential is used as the confining potential, the location of its minimum indicating the location of the interface. This potential is applied to all the LJ beads that a composite particle is made of. Specifically, an LJ bead experiences a force pointing toward the suspension phase when it intersects with the liquid-vapor interface, as in
\begin{equation}
    \label{eqn:mim_f}
    F^{N}_{i}=k_{s}(\Delta z_{i}+R\cos\theta) \text{ for } |\Delta z_{i}| \le R~,
\end{equation}
where $k_{s}=3.0\epsilon/\sigma^{2}$ is a spring constant, $R=1.0 \sigma$ is the nominal size of the LJ bead, $\theta$ is a contact angle capturing the interaction between the LJ bead and the solvent. In this study, $\theta = 0$ is adopted to make the LJ beads solvophilic, which ensures that all the composite particles remain within the suspension phase. Since the liquid-vapor interface simulated here is parallel with the $xy$ plane, the remaining variable is $\Delta z_{i} \equiv z_b - z_i$, where $z_b$ is the coordinate of the LJ bead under consideration along the $z$ axis and $z_i$ is the location of the liquid-vapor interface, that is, the minimum location of the confining potential.

The liquid-vapor interface, implemented via the confining potential, is initially placed at $z=800\sigma$. The evaporation process is initiated after the suspension is well equilibrated. During evaporation, the location of the liquid-vapor interface is moved along the $-z$ direction at a speed of $v_e$, by moving the minimum location of the confining potential. Therefore, $z_i$ changes with time as
\begin{equation}\label{eqn:mim}
     z_i(t)=z_i(0) -v_{e} \Delta t~,
\end{equation}
where $z_i(0) = 800\sigma$ and $\Delta t$ is the time elapsed after the start of the evaporation process.

In the following, we report results for two different evaporation rates, one faster with $v_{e} = 1.311\times 10^{-2}\sigma/\tau$ and another slowed by a factor of 10 with $v_{e} = 1.311\times 10^{-3}\sigma/\tau$. The effect of evaporation rate on particle distribution in a drying film is often characterized through a P\'{e}clet number, defined as $\text{Pe} = Hv_e/D$, where $H$ is the thickness of the film and $D$ is the diffusion coefficient of the particle. The P\'{e}clet number represents the competition between the evaporation-induced advection of a particle and its diffusion during a drying process. Since the film shrinks as evaporation progresses, the initial thickness of the film is usually used to compute Pe. The diffusion coefficient of each type of composite particles in a mixture is computed with independent MD simulations of the same mixture dispersed in a cubic box of side length of $200\sigma$, with periodic boundary conditions applied to the three Cartesian directions. The volume fraction of each type of particles is the same as their initial volume fraction in the evaporating systems. The same Langevin thermostat is applied to each LJ bead as the one used in the equilibration and evaporation simulations.

The results on the diffusion coefficient ($D$) and P\'{e}clet number are all summarized in Table~\ref{table:systems}. It can be noted that $D$ is inversely proportional to particle mass. This is because each LJ bead is coupled to the implicit solvent through the same Langevin thermostat. That is, it is assumed that the solvent drains freely through the composite particles as they move and each LJ bead experiences the same drag coefficient, $m/\Gamma$, where $m$ is the mass of the LJ bead and $\Gamma$ is the damping time. Assuming the Einstein relation, the diffusion coefficient of a composite particle can be computed from $D=\Gamma k_\text{B}T/(Nm)$, where $N$ is the number of LJ beads it contains. Clearly, the diffusion coefficient is inversely proportional to $Nm$, the particle mass. In this sense, the computational model adopted here for composite particles mimics the Rouse model of polymers.\cite{RubinsteinColbyBook} For the solid sphere containing 527 LJ beads, its diffusion coefficient is estimated to be $\sim 3\times 10^{-2}\sigma^2/\tau$ for $\Gamma = 15.7\tau$, which is close to the value determined with independent MD simulations computing the mean-square displacement of the particles in each mixture at equilibrium (see Supplementary Material). For reasons explained later, Table~\ref{table:systems} also includes the surface area and sphericity of each type of composite particles. These parameters capture the main attributes of the shape of particles that affect their distribution in a drying film.

\section{Results and Discussion}\label{sec:res}

\begin{figure*}
    \centering
    \includegraphics[width=0.9\textwidth]{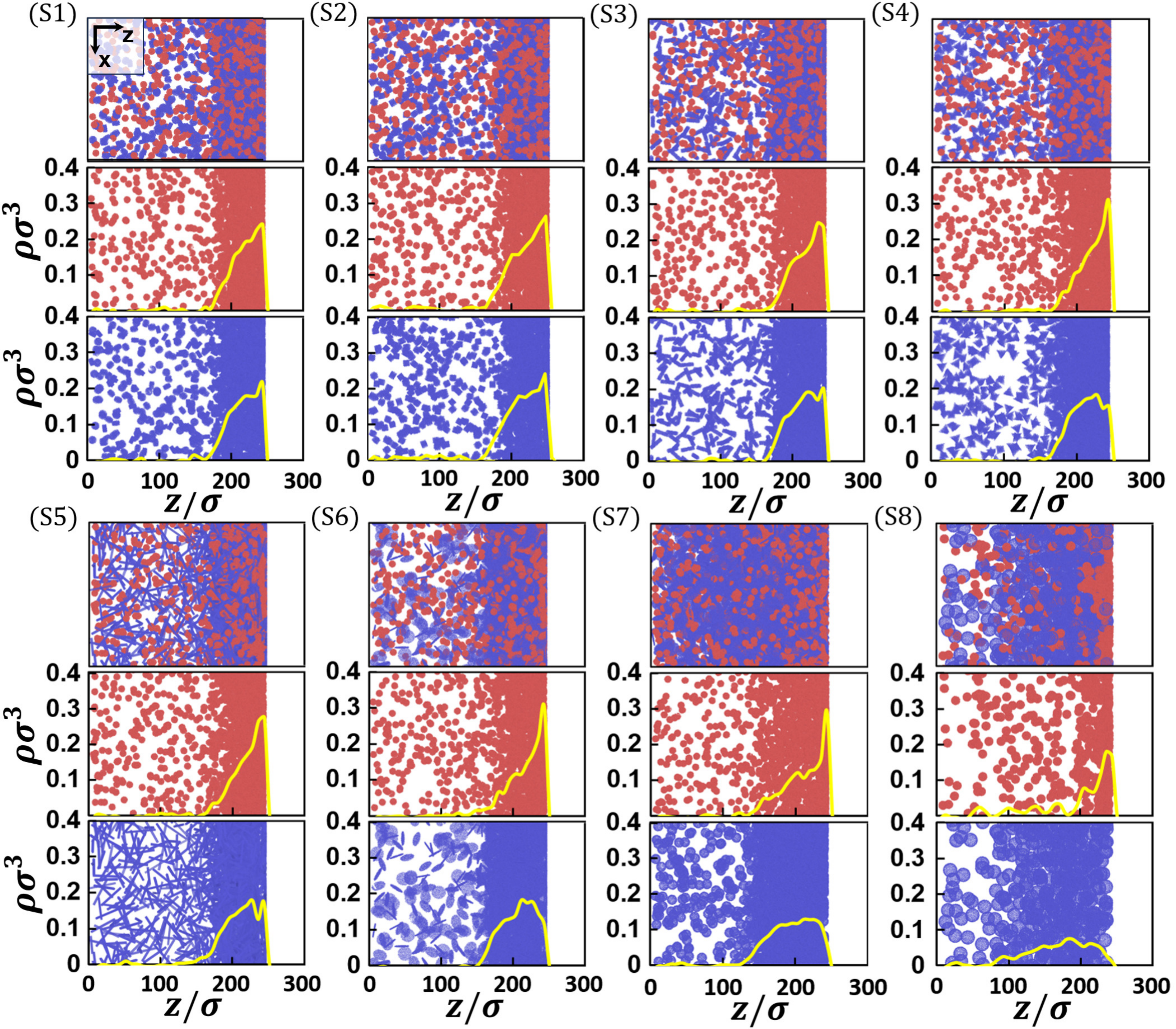}
    \caption{Projections in the $xz$ plane of particle distribution in the drying film of a thickness of $\sim 240\sigma$ for systems S1-S8 in Table~\ref{table:systems} under the \textbf{fast-drying} condition with $v_{e} = 1.311\times 10^{-2}\sigma/\tau$. Each subset contains three images: mixture (top), solid spheres (middle), and aspherical/hollow particles (bottom). The yellow lines indicate the number density of the corresponding type of particles in the film along the direction of drying.}
    \label{fig:snapshot_film_240sigma}
\end{figure*}

Figure~\ref{fig:snapshot_film_240sigma} shows the distribution of particles in the evaporating films when their thickness decreases to $\sim 240\sigma$, which is 30\% of the initial thickness. The data presented are collected under the fast-drying condition with $v_{e} = 1.311\times 10^{-2}\sigma/\tau$. Snapshots showing the evolution of particle distribution during drying as well as a plot corresponding to Fig.~\ref{fig:snapshot_film_240sigma} but under the slow-drying condition ($v_{e} = 1.311\times 10^{-3}\sigma/\tau$) are included in the Supplementary Material. In each system in Fig.~\ref{fig:snapshot_film_240sigma}, there is a clear accumulation of particles near the top of the drying film, indicating that all the simulations reported here are in the rapid evaporation regime with $\text{Pe} \gg 1$.

To quantify the distribution of particles, the number density of each type of composite particles along the drying direction is calculated as $\rho(z)$, which is the number of LJ beads, the building block of the composite particles, in a spatial bin from $z-1.0\sigma$ to $z+1.0\sigma$ divided by the volume of the bin. The profiles of $\rho(z)$ are included in Fig.~\ref{fig:snapshot_film_240sigma} as yellow lines. In all systems, the number density of the solid spheres increases monotonically as the liquid-vapor interface is approached from the suspension side. However, the distribution of aspherical particles depends on their sphericity, and thus the surface area. For aspherical particles with high sphericity, including thick disks, cubes, and short rods, the number density increases monotonically from its bulk value and peaks near the liquid-vapor interface. A careful examination of the snapshots on the particle distribution in the corresponding drying films confirms that stratification does not occur in systems S1, S2, and S3 (see the Supplementary Material). However, for aspherical particles with lower sphericity, including regular tetrahedra, long rods, and thin disks, the number density increases first and then peaks at a location well below the liquid-vapor interface. Then their number density starts to decrease toward the interface, indicating the possibility of stratification if we adopt the criterion used in a previous study by Zhou \textit{et al.}\cite{Zhou2017} This observation is further corroborated with a detailed analysis of the evolution of particle distribution in the drying films, as discussed below.

Stratification in a drying film containing a dispersed mixture of colloidal particles is typically understood on the basis of the P\'{e}clet number and diffusiophoresis.\cite{Fortini2016, Zhou2017, Sear2017, Sear2018, Tang2018Langmuir, Schulz2018} With a binary mixture of large and small solid spheres as an example, the key parameter is the particle size ratio, $R_L/R_S$. If the Stokes-Einstein relationship holds, then the ratio between the diffusion coefficients of the two types of particles is just the reciprocal of the size ratio. As a result, the P\'{e}clet number of the large particles is larger than that of the small particles by the size ratio. That is, $\text{Pe}_L/\text{Pe}_S = R_L/R_S > 1$. In the case of rapid evaporation where $\text{Pe}_L > \text{Pe}_S \gg 1$, both types of particles are first enriched near the evaporation front when evaporation starts, resulting in concentration gradients. Then diffusiophoresis begins to operate, where particles migrate in response to the concentration gradient of other particles, but the effects are not symmetric between particles with different sizes. In hard-sphere-like systems, the mobility of large particles induced by the concentration gradient of small particles is greater than that of small particles driven by the concentration gradient of large particles.\cite{Zhou2017} As a result, large particles are pushed out of the region near the evaporation front, leaving small particles there and creating the so-called ``small-on-top'' stratification.\cite{Fortini2016}

The physical picture presented above provides a basis for understanding stratification phenomena in drying films containing solid spherical particles. However, the results in Fig.~\ref{fig:snapshot_film_240sigma} indicate that the shape of the particles can also play an important role in determining their distribution in a drying film. Specifically, the results seem to follow the order ranked by the sphericity of the particles. For the aspherical particles with higher sphericity and thus lower surface area (in systems S1, S2, and S3), their number density in the drying film increases monotonically toward the receding liquid-vapor interface, similar to the number density of the solid spheres. These systems are unstratified after drying. However, for the aspherical particles with lower sphericity (in systems S4, S5, and S6), their number density peaks at a location away from the receding interface. Thus, systems S4 to S6 can be classified as stratified. Furthermore, it is noted that as the sphericity of the aspherical particles decreases, the signature of stratification is enhanced, reflected by the peak location of the number density of the aspherical particles further shifting away from the receding interface. By definition, the sphericity of a particle is inversely proportional to its surface area when its volume (i.e., mass for nonhollow particles) is fixed. Therefore, the results in Fig.~\ref{fig:snapshot_film_240sigma} point to the importance of the surface area of the aspherical particles in determining their distribution with respect to the solid spheres, although the two have similar masses and diffusion coefficients.

\begin{figure*}
    \centering
    \includegraphics[width=0.9\textwidth]{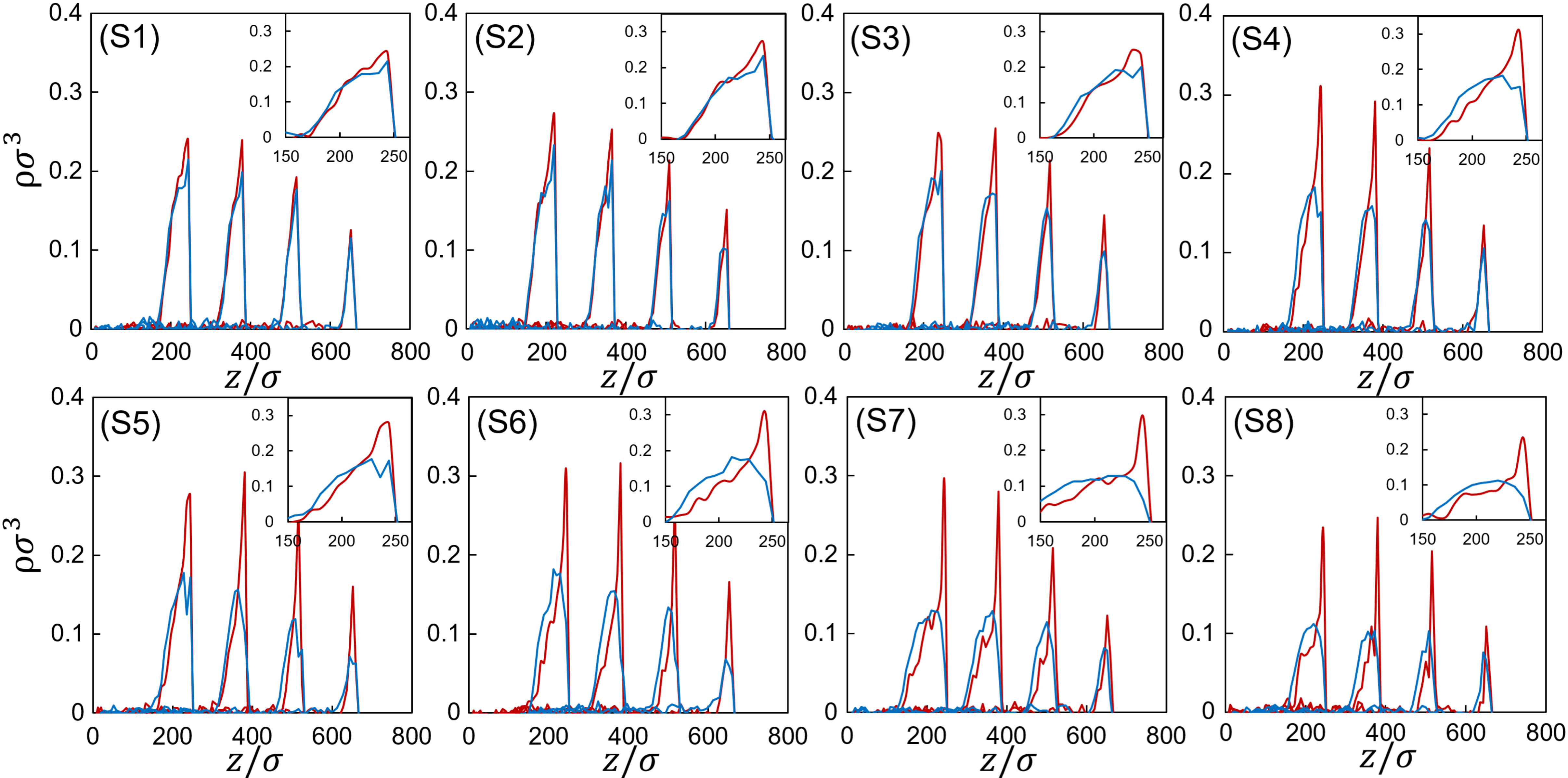}
    \caption{Evolution of particle distribution in the drying films: solid spheres (red lines) and aspherical/hollow particles (dark blue lines). The inset shows the particle distribution when the thickness of the corresponding film is reduced to $\sim 240\sigma$. The data are for the \textbf{fast-drying} condition with $v_{e} = 1.311\times 10^{-2}\sigma/\tau$.}
    \label{fig:den_profile_fast_drying}
\end{figure*}

To further test the hypothesis that the surface area ratio of particles is in fact a key parameter controlling stratification induced by evaporation in particle mixtures, we turn to system S7, which contains a mixture of solid and hollow spheres with similar masses but different surface areas. Technically speaking, both particles have a sphericity of 1. They also have similar diffusion coefficients as their masses are close. However, due to the contrast in their surface areas (with a ratio of about $1.96$), the mixture is clearly stratified after drying with the solid spheres enriched at the evaporation front, as shown in the result for S7 in Fig.~\ref{fig:snapshot_film_240sigma}. As a further validation, we simulate a mixture of solid and hollow spheres with larger sizes but still similar masses, labeled system S8 in Table~\ref{table:systems}. The surface area ratio for S8 is about $2.37$. Since the particle mass is more than double that in system S7, the particle diffusion coefficients are reduced in S8, leading to larger P\'{e}clet numbers and placing S8 in a regime of more rapid evaporation under the same $v_e = 1.311\times 10^{-2}\sigma/\tau$. The results in Fig.~\ref{fig:snapshot_film_240sigma} show that S8 is more stratified than S7, indicating that stratification in a mixture of particles with similar masses is enhanced when the contrast of their surface areas and/or evaporation rates are increased. Therefore, the surface area ratio is another key parameter to consider, in addition to the ratio of diffusion coefficients (which might be translated into the size ratio for solid spheres as in most studies reported in the literature) and the evaporation speed, when we are concerned with drying films of particle mixtures and the possibility of stratified particle distribution after solvent evaporation.

The evolution of particle distribution in a drying film can be revealed by examining the density profile of the particles at various stages of drying. Such results for all systems in Table~\ref{table:systems} under the fast-drying condition ($v_{e} = 1.311\times 10^{-2}\sigma/\tau$) are included in Fig.~\ref{fig:den_profile_fast_drying}, with the inset in each subfigure showing the particle distribution when the film thickness is reduced to $\sim 240 \sigma$. In each equilibrium film prior to evaporation, the particles are uniformly dispersed. After evaporation begins, both solid spheres and other particles quickly accumulate at the evaporation front. For systems S1 and S2, the density profiles of the two types of particles essentially trace each other as evaporation proceeds, with the peak density of the solid spheres slightly higher than that of the aspherical particles (thick disks in S1 and cubes in S2). The difference in the peak density of the two types of particles increases slightly in system S3, in which the solid spheres and the short rods (see Fig.~\ref{fig:comp_particle_shape}) are mixed. Although in system S3 the location of the peak density largely overlaps for the two types of particles, this system exhibits a signature consistent with crossing over from an unstratified state to a stratified state.

The density profiles of systems S4, S5, and S6 reveal the clear onset of stratification. For these systems in the early stage of drying, both types of particles accumulate at the evaporation front, where their density peaks. As evaporation proceeds, the peak density value of the solid spheres increases with time, but its location coincides with the receding evaporation front. At the same time, the density peak of the aspherical particles starts to shift away from the receding liquid-vapor interface, and this trend is stronger for systems containing particles with lower sphericity values and thus larger surface areas. When the film thickness is reduced to $\sim$30\% of its initial value, the particles clearly show a stratified distribution, with a surface layer at the evaporation front enriched with the solid spheres. Below this surface layer, there is a zone in which the aspherical particles have a higher density.

The degree of stratification is further enhanced in systems S7 and S8, which consist of solid and hollow spheres. In these systems, the peak locations of the density profiles of the particles are separated even in the early stage of evaporation. The contrast between the peak values for the two types of particles in the mixture also grows more pronouncedly with time. The density profiles in the films of thickness $240\sigma$ clearly show a stratified distribution of the particles. Since the solid and hollow spheres in S7 or S8 have similar masses, their diffusion coefficients are close, and the contrast in their diffusion rates is insufficient to drive stratification. The results thus indicate that the observed stratified particle distribution is due to the difference in surface area between the particles. Comparing the results for S7 and S8 also indicates that as the surface area ratio between the two types of particles in the mixture increases, the extent of stratification increases.

The implications of the results shown in Fig.~\ref{fig:den_profile_fast_drying} can be summarized as follows. In a mixture of spheres and aspherical particles of similar masses (to produce close diffusion coefficients), stratification can occur if the sphericity of the latter is so low that they differ significantly from the spheres in terms of surface area. In this sense, the shape of a particle takes effect through its surface area. Since in the computational model adopted here, particles with similar masses have close diffusion coefficients, a plausible hypothesis is that the surface area of a particle influences its diffusiophoretic mobility in a mixture featuring shape dispersity. That is, there is an asymmetry between particles of various shapes at the same mass; particles with larger surface areas are more strongly driven by the density gradient of particles with smaller surface areas. A larger ratio of surface area leads to a stronger asymmetry. This is analogous to the situation of mixing solid spheres with size dispersity, where cross interaction has a greater effect on the diffusiophoretic motion of larger particles.\cite{Zhou2017} Since larger particles have larger surface areas, the effects of shape and size dispersity could be unified into the same framework based on the notion of surface area dispersity, where the diffusiophoretic mobility of a particle is positively correlated with its surface area. This correlation may be more easily validated if the diffusion coefficient of the particle is kept (nearly) unchanged when its surface area is varied by changing its shape. This hypothesis remains to be tested with more studies in the future.

The importance of surface area can be justified by noting that the excluded volume of particles is an essential factor in determining the free energy of a particle mixture.\cite{Lekkerkerker2011Book} For aspherical particles, their excluded volume depends not only on the physical volumes of the particles involved but also on their surface areas. Consider hard rods of cross-sectional diameter $d$ and length $l$, with $l \ge d$. Its physical volume is $V_p = \pi d^2 l/4$. The excluded volume between a pair of such rods is $V_e = 2 l^2 d |\sin \alpha |$, where $\alpha$ is the angle between the rods.\cite{Doi2013Book} The average excluded volume is therefore $\langle V_e \rangle = 4 l^2 d / \pi$. Clearly, $\langle V_e \rangle$ cannot be determined from $V_p$ alone but can be calculated if the surface area of the rod, $A_p = \pi d^2/2+ \pi d l$, is also known. It is further noted that for a given $V_p$, the surface area $A_p$ is minimized at $l=d$. This is not surprising, as a rod with $l=d$ has an aspect ratio of 1 and is the closest to a sphere among all rods. Furthermore, the average excluded volume can be written as $\langle V_e \rangle = \frac{8 V_p}{\pi}\frac{l}{d}$. When the aspect ratio of the rod increases under the constraint of a constant $V_p$, the diameter of the rod $d$ decreases and the length $l$ increases. As a result, both $A_p$ and $\langle V_e \rangle$ increase. Similar arguments can also be made for rods with $l < d$, which are termed disks in this paper, and other aspherical particles. For a pair of solid/hollow spheres with outer radii $r_1$ and $r_2$, the excluded volume is $V_e = \frac{2\pi}{3}\left( r_1 + r_2 \right)^2$. Therefore, for spheres, the larger the outer size, the larger the surface area and the excluded volume. In general, when the physical volumes of hard particles are kept constant, the average excluded volume between the particles increases with increasing surface areas (i.e., decreasing sphericity). The driving force behind diffusiophoresis in a particle mixture is the gradient in free energy (or, more accurately, the chemical potential) which depends on the excluded volumes of the particles in the mixture.\cite{Zhou2017} Therefore, the surface area of a particle plays a crucial role in determining its diffusiophoretic mobility in a mixture. In general, particles with larger surface areas are expected to exhibit stronger diffusiophoretic responses because their average excluded volumes are larger.

\begin{figure*}
    \centering
    \includegraphics[width=0.9\textwidth]{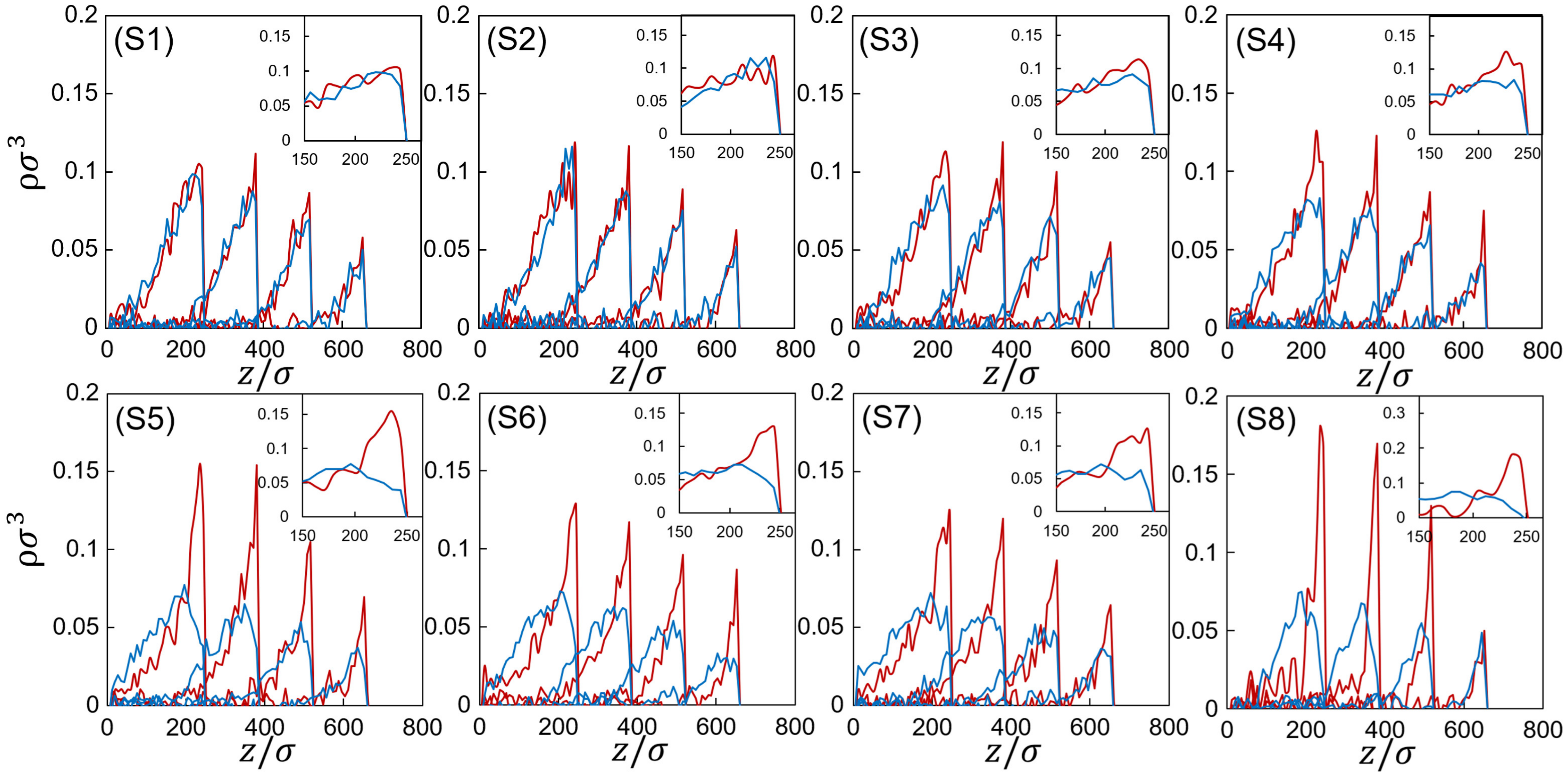}
    \caption{Evolution of particle distribution in the drying films: solid spheres (red lines) and aspherical/hollow particles (dark blue lines). The inset shows the particle distribution when the thickness of the corresponding film is reduced to $\sim 240\sigma$. The data are for the \textbf{slow-drying} condition with $v_{e} = 1.311\times 10^{-3}\sigma/\tau$.}
    \label{fig:den_profile_slow_drying}
\end{figure*}

The effect of evaporation speed on drying outcome is shown in Fig.~\ref{fig:den_profile_slow_drying}, which includes results at an evaporation rate 10 times lower than previous, now at $v_{e} = 1.311\times 10^{-3}\sigma/\tau$. Again, stratification does not occur in systems S1 and S2, while S3 appears to be at the crossover into the stratification regime. Systems S4 to S8 all show clear signs of stratification. Although the peak values of the particle density decrease with the slower evaporation rate, there is a thickening of the surface layer, where the solid spheres exhibit a density more than twice that of the aspherical or hollow particles. This thickening is consistent with the widening of the zone affected by the receding liquid-vapor interface, where the particle density clearly deviates from the corresponding bulk value. The thickening of the stratified surface layer in Fig.~\ref{fig:den_profile_slow_drying} indicates that the extent of stratification can be enhanced with slower evaporation. Of course, if evaporation is too slow, then no stratification is expected. The trend thus echos a previous finding that stratification is more pronounced at an intermediate evaporation rate.\cite{Tang2019JCP_compare}

\begin{figure}[hbtp] 
    \centering
    \includegraphics[width=0.4\textwidth]{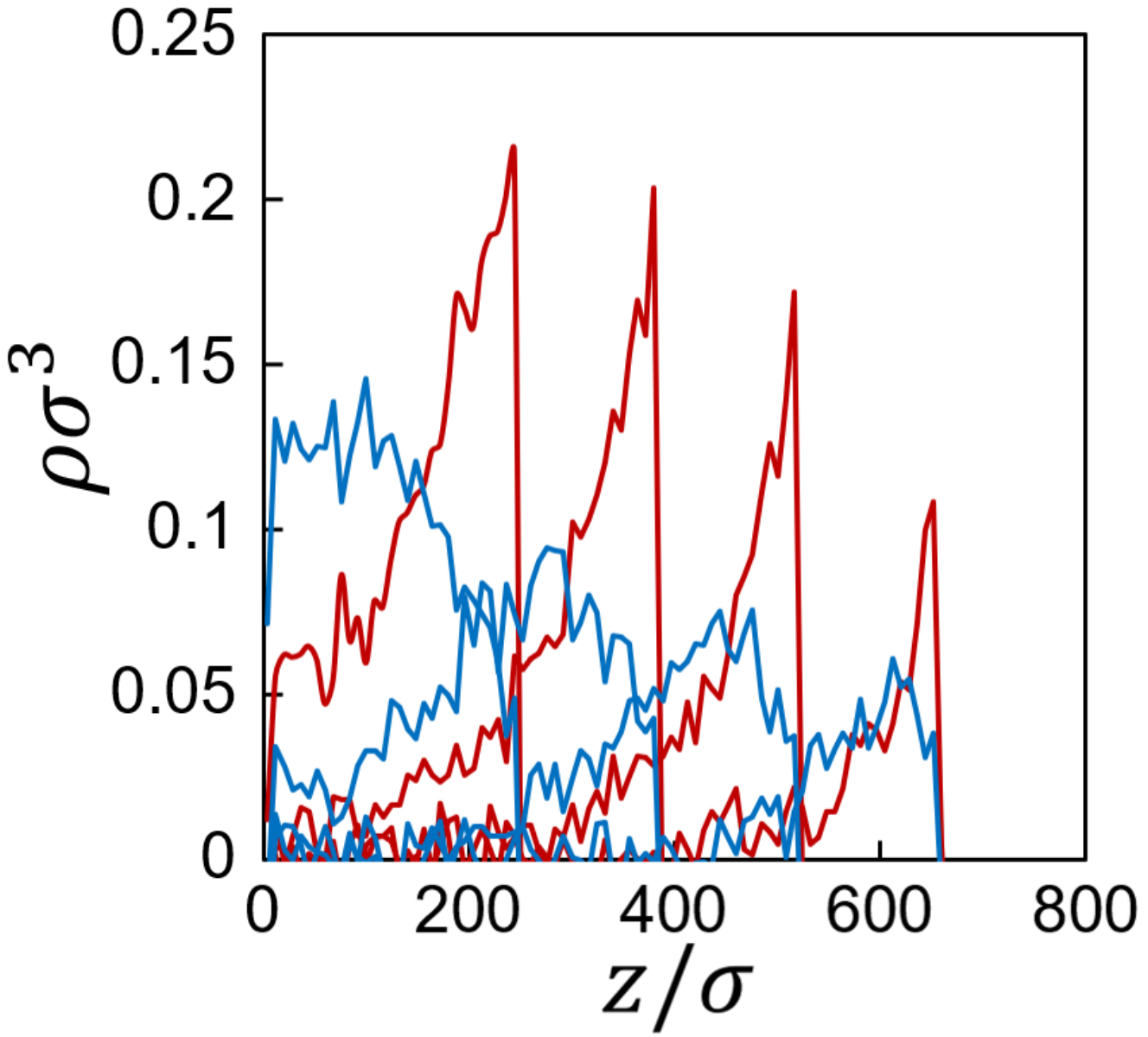}
    \caption{Evolution of particle distribution in the drying film: solid spheres (red lines) and hollow spheres (dark blue lines). Both have the same outer radius of $5\sigma$. The data are for the \textbf{slow-drying} condition with $v_{e} = 1.311\times 10^{-3}\sigma/\tau$.}
    \label{fig:density_same_size_spheres}
\end{figure}

The results presented so far indicate that under a given concentration gradient of other particles, the diffusiophoretic mobility of a particle depends on its surface area (or excluded volume). Therefore, in a mixture of solid and hollow spheres with the same outer radius and thus the same surface area, the diffusiophoretic contrast can be removed and the particle distribution after drying is expected to be determined solely by the difference in their diffusion coefficients. The evolution of particle distribution for one such system under the slow-drying condition is presented in Fig.~\ref{fig:density_same_size_spheres}, where both solid and hollow spheres have the same radius of $5\sigma$. The inner radius of the hollow sphere is $4\sigma$, making its mass only $\sim 50\%$ of that of the solid sphere. As a result, the solid spheres diffuse more slowly with a diffusion coefficient of $0.0244\sigma^2/\tau$, which is about half the value of the hollow spheres, $0.0485\sigma^2/\tau$. As expected, the solid spheres are significantly enriched at the evaporation front after drying, leading to clear ``slower-on-top'' stratification. This is consistent with the expectation that diffusion favors the accumulation of larger particles at the top of the drying film in the case of a bidisperse mixture of small and large solid spheres. In that case, the large particles are the species that diffuses more slowly. However, in bidisperse solid sphere mixtures, the larger spheres also have a larger surface area, and thus larger diffusiophoretic mobility, which favors their motion leaving the evaporation front. At fast evaporation ($\text{Pe}_L > \text{Pe}_S \gg 1$), diffusiophoresis wins the race and leads to the ``small-on-top'' stratification. Only when $\text{Pe}_L > 1 > \text{Pe}_S$ do we expect ``large-on-top'' (that is, ``slower-on-top''). However, the corresponding window of parameter space is limited and hard to realize experimentally.

\begin{figure}[htbp] 
    \centering
    \includegraphics[width=0.4\textwidth]{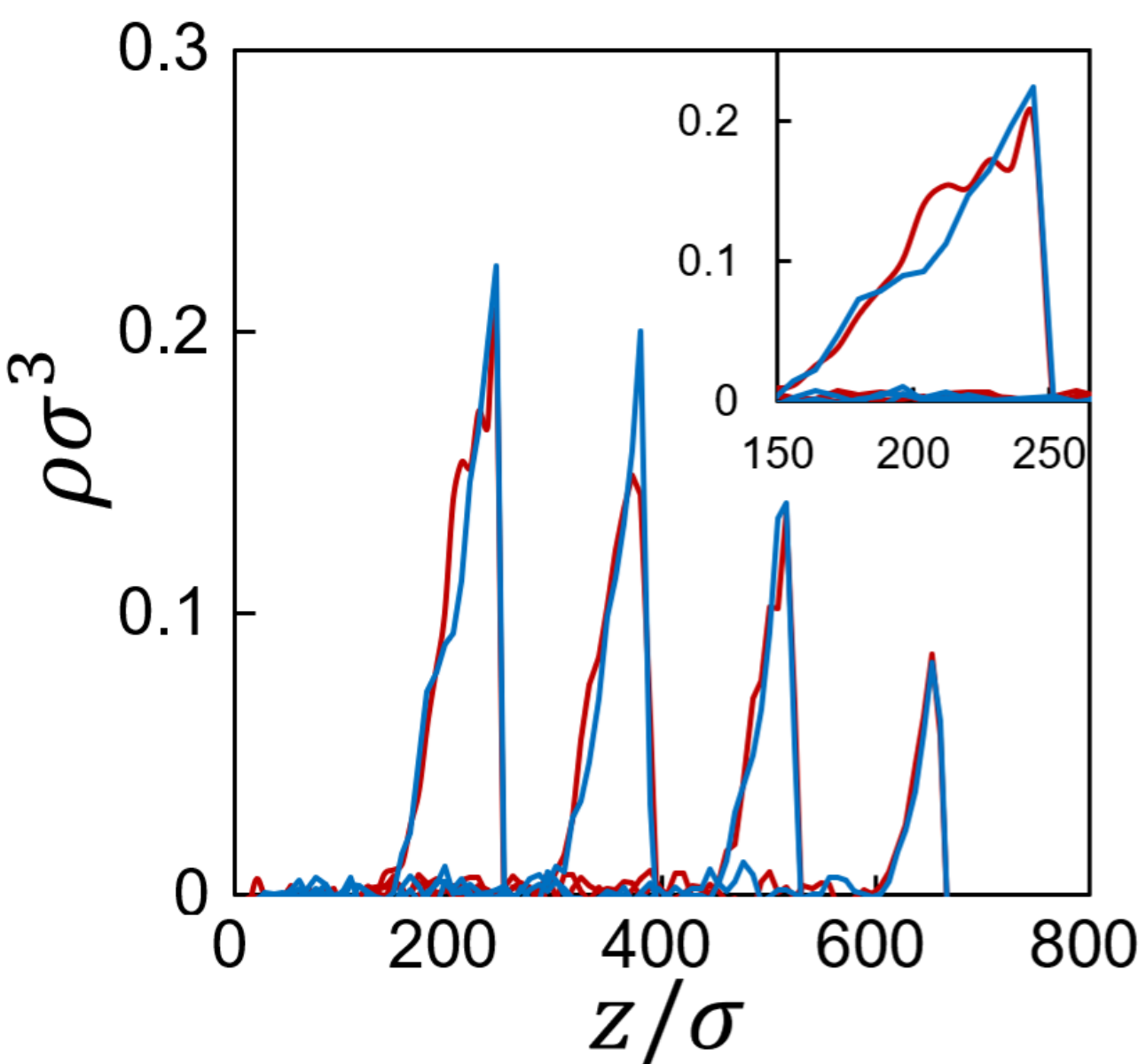}
    \caption{Evolution of particle distribution in the drying film: thin disks (red lines) and long rods (dark blue lines). The two have very different shapes but their surface areas differ only by $\sim 10\%$ (see Table~\ref{table:systems}). The data are for the \textbf{fast-drying} condition with $v_{e} = 1.311\times 10^{-2}\sigma/\tau$.}
    \label{fig:density_thin_disk_long_rod}
\end{figure}

To further test the hypothesis that the difference in surface area is an appropriate measure of the shape asymmetry between particles, a mixture of the thin disks and the long rods in Fig.~\ref{fig:comp_particle_shape} is examined, with the initial volume fraction of each set at 0.02. The two are apparently distinct in shape with contrasting aspect ratios, 0.084 for the thin disk and 20 for the long rod. However, the surface areas of the two differ only by $\sim 10\%$ and they are expected to exhibit similar diffusiophoretic responses. Furthermore, the two have similar diffusion coefficients, and the contrast of their diffusion behavior is expected to be marginal. Taking all these factors into account, the expectation is that this mixture should not stratify during drying, which is confirmed by the results in Fig.~\ref{fig:density_thin_disk_long_rod}. The density profiles of the two types of particles essentially trace each other, with the peak value slightly higher for the long rods. Even this minor difference is consistent with the observation that the long rod has a slightly smaller surface area than the thick disk and is thus expected to have a slightly weaker diffusiophoretic response, which enhances their density at the evaporation front compared to the thick disks.

\section{Conclusions}\label{sec:conc}

Using large-scale molecular dynamics simulations, here we show that, similar to previous studies of disperse spheres, stratification can occur in the presence of shape dispersity. Specifically, in a suspension film containing a binary mixture of solid spheres and aspherical particles with similar masses (and thus close diffusion coefficients as implemented in our computational model), stratification occurs after drying when the sphericity of the aspherical particles is so low that their surface area (and excluded volume) is significantly larger than that of the solid spheres. In these systems, the solid spheres are always enriched in the top region of the stratified dry films. Such stratification also occurs in a suspension film in which the solid spheres are mixed with hollow spheres with similar masses (and thus close diffusion coefficients). In this case, the hollow spheres have a larger surface area than the solid spheres and are pushed away from the evaporation front, leading to a stratified distribution with the solid spheres enriched at the evaporation front. Furthermore, when the surface area of the aspherical particles or the hollow spheres is not that different from the surface area of the solid spheres, stratification is not observed after drying. In a mixture of aspherical particles (e.g., long rods and thin disks) with apparently distinct shapes but similar surface areas and diffusion coefficients, stratification does not occur either.

The role of surface area revealed in this study may provide another perspective to understand the phenomenon of drying-induced stratification in colloidal films that feature size dispersity.\cite{Fortini2016, Zhou2017, Howard2017, Tang2018Langmuir} For solid spheres with different radii, the larger spheres diffuse more slowly, which promotes their accumulation at the evaporation front during fast drying. However, the larger spheres also have larger surface areas (and excluded volumes), which enhance their diffusiophoretic mobilities. The net result is that in the regime where evaporation is sufficiently fast even for the smaller spheres, diffusiophoresis dominates and the larger spheres are driven out of the region near the evaporation front, leaving the smaller spheres there and resulting in ``small-on-top'' stratification. If the larger spheres are modified in such a way that their surface areas do not change but their diffusion coefficients become close to those of the smaller spheres, then the diffusion contrast, which favors ``slower-on-top'' (i.e., ``large-on-top'' for solid spheres), is removed and stronger ``small-on-top'' stratification is expected. This expectation is confirmed by the results presented here for systems S7 and S8 (see Figs.~\ref{fig:den_profile_fast_drying} and \ref{fig:den_profile_slow_drying}). On the other hand, if two types of particles have close surface areas but distinctive diffusion coefficients, then ``slower-on-top'' stratification will emerge under fast evaporation conditions, where ``fast'' implies that the corresponding evaporation rates lead to sufficiently large P\'{e}clet number even for the more diffusive particles in the mixture.

\section*{Acknowledgments}

This material is based on work supported by the National Science Foundation under Grant No. DMR-1944887. This research was initially supported by a 4-VA Collaborative Research Grant (``Material Fabrication via Droplet Drying''). This work was performed, in part, at the Center for Integrated Nanotechnologies, an Office of Science User Facility operated for the U.S. Department of Energy Office of Science. Sandia National Laboratories is a multimission laboratory managed and operated by National Technology and Engineering Solutions of Sandia, LLC., a wholly owned subsidiary of Honeywell International, Inc., for the U.S. Department of Energy's National Nuclear Security Administration under contract DE-NA0003525. This paper describes objective technical results and analysis. Any subjective views or opinions that might be expressed in the paper do not necessarily represent the views of the U.S. Department of Energy or the United States Government.


\clearpage
\newpage
\onecolumngrid
\renewcommand{\thesection}{S\arabic{section}}
\setcounter{section}{0}
\renewcommand{\thefigure}{S\arabic{figure}}
\setcounter{figure}{0}
\renewcommand{\theequation}{S\arabic{equation}}
\setcounter{equation}{0}
\renewcommand{\thepage}{SI-\arabic{page}}
\setcounter{page}{1}
\begin{center}
{\bf \large Supplementary Information for ``Effect of Particle Shape on Stratification in Drying Films of Binary Colloidal Mixtures''}
\end{center}
\begin{center}
{\bf Binghan Liu$^{1}$, Gary S. Grest$^{2}$, and Shengfeng Cheng$^{1,3}$}
\end{center}
\begin{center}
{$^1$Department of Physics, Center for Soft Matter and Biological Physics, and Macromolecules Innovation Institute, Virginia Tech, Blacksburg, Virginia 24061, USA\\$^2$Sandia National Laboratories, Albuquerque, NM 87185, USA\\ $^3$Department of Mechanical Engineering, Virginia Tech, Blacksburg, Virginia 24061, USA}
\end{center}

\section{Snapshots of Drying Films}

In this section, we include a series of snapshots of the drying film for each system, illustrating the evolution of the particle distribution during drying. In each set of images, the left column is for the fast-drying condition with $v_{e} = 1.311\times 10^{-2}\sigma/\tau$, while the right column is for the same system under slow-drying with $v_{e} = 1.311\times 10^{-3}\sigma/\tau$. In Fig.~\ref{fg:snapshots_sys_9_slow} for system S9, snapshots are shown only for the slow-drying condition, while in Fig.~\ref{fg:snapshots_sys_10_fast} for the mixture of thin disks and long rods, snapshots are included only for the fast-drying condition. In these images, the solid spheres are shown in red, while the particles of the other type (aspherical or hollow) are shown in blue. In Fig.~\ref{fg:snapshots_sys_10_fast}, the thin disks are shown in red, while the long rods are colored blue. The various particle shapes studied here are summarized in Figure 1 of the main text.

\begin{figure}[htb]
\includegraphics[width = 0.8\textwidth]{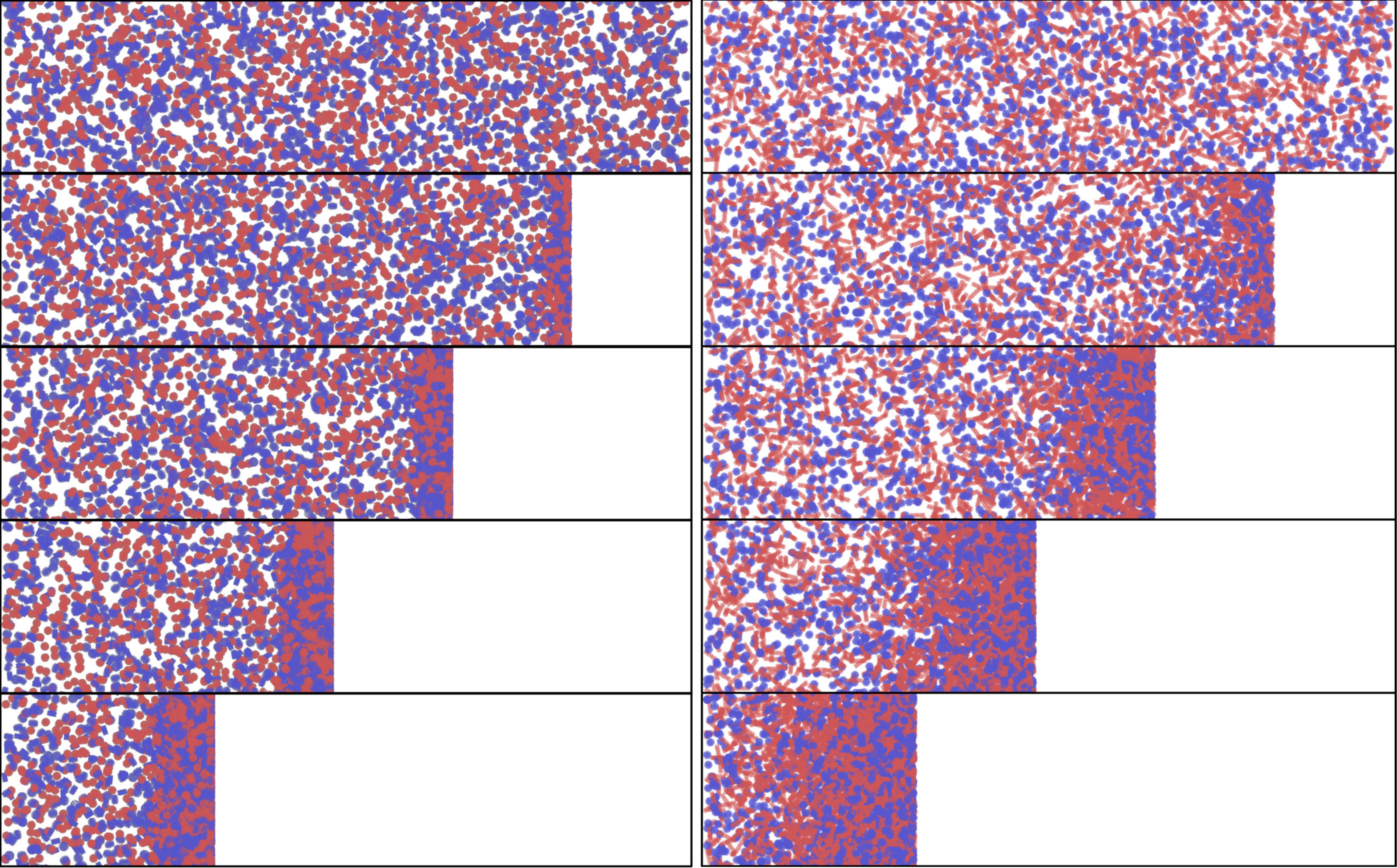}
\caption{Snapshots of the drying film for system S1 (solid spheres + thick disks).}
\label{fg:snapshots_sys_1}
\end{figure}

\begin{figure}[htb]
\includegraphics[width = 0.8\textwidth]{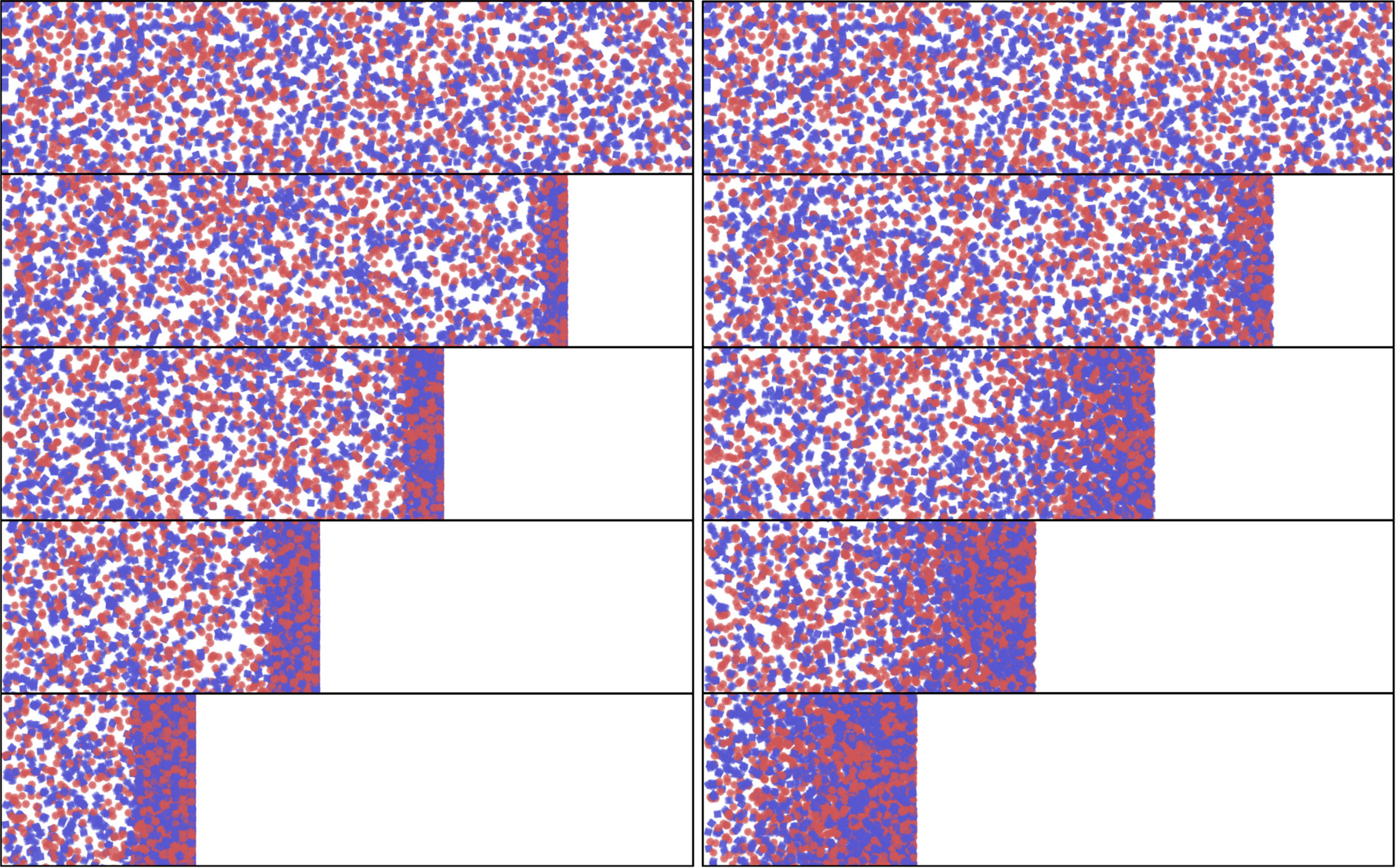}
\caption{Snapshots of the drying film for system S2 (solid spheres + cubes).}
\label{fg:snapshots_sys_2}
\end{figure}

\begin{figure}[htb]
\includegraphics[width = 0.8\textwidth]{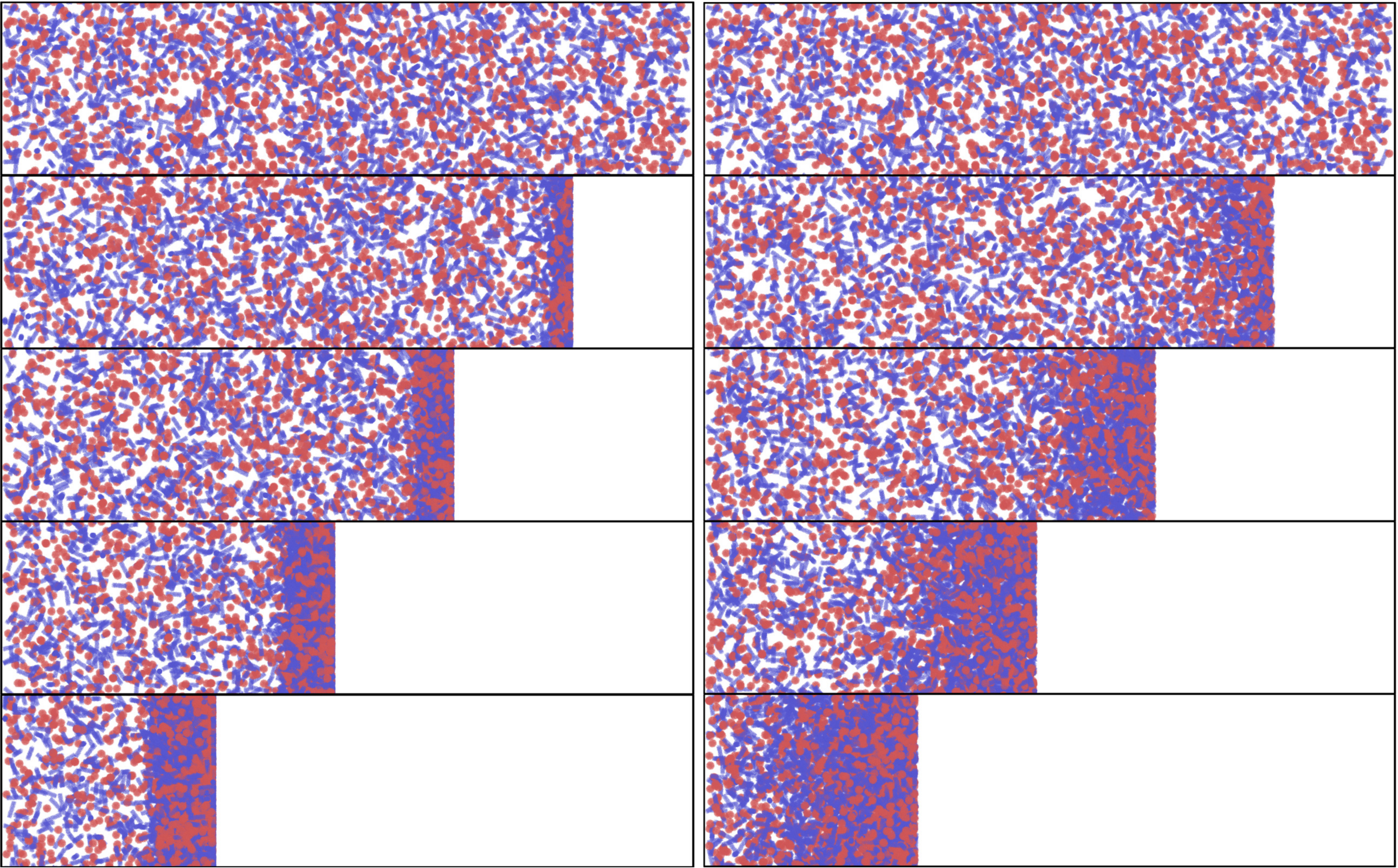}
\caption{Snapshots of the drying film for system S3 (solid spheres + short rods).}
\label{fg:snapshots_sys_3}
\end{figure}

\begin{figure}[htb]
\includegraphics[width = 0.8\textwidth]{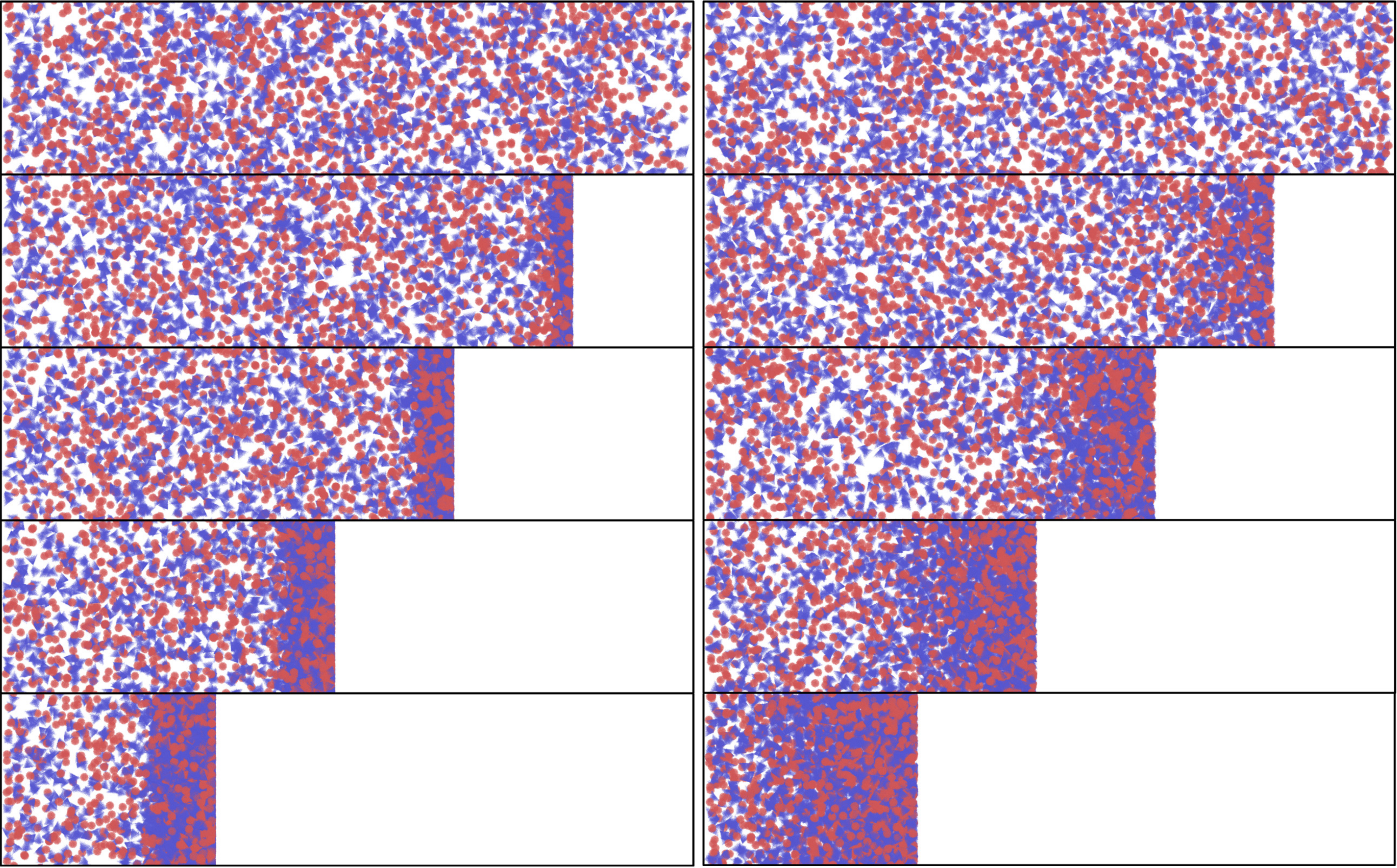}
\caption{Snapshots of the drying film for for system S4 (solid spheres + tetrahedra).}
\label{fg:snapshots_sys_4}
\end{figure}

\begin{figure}[htb]
\includegraphics[width = 0.8\textwidth]{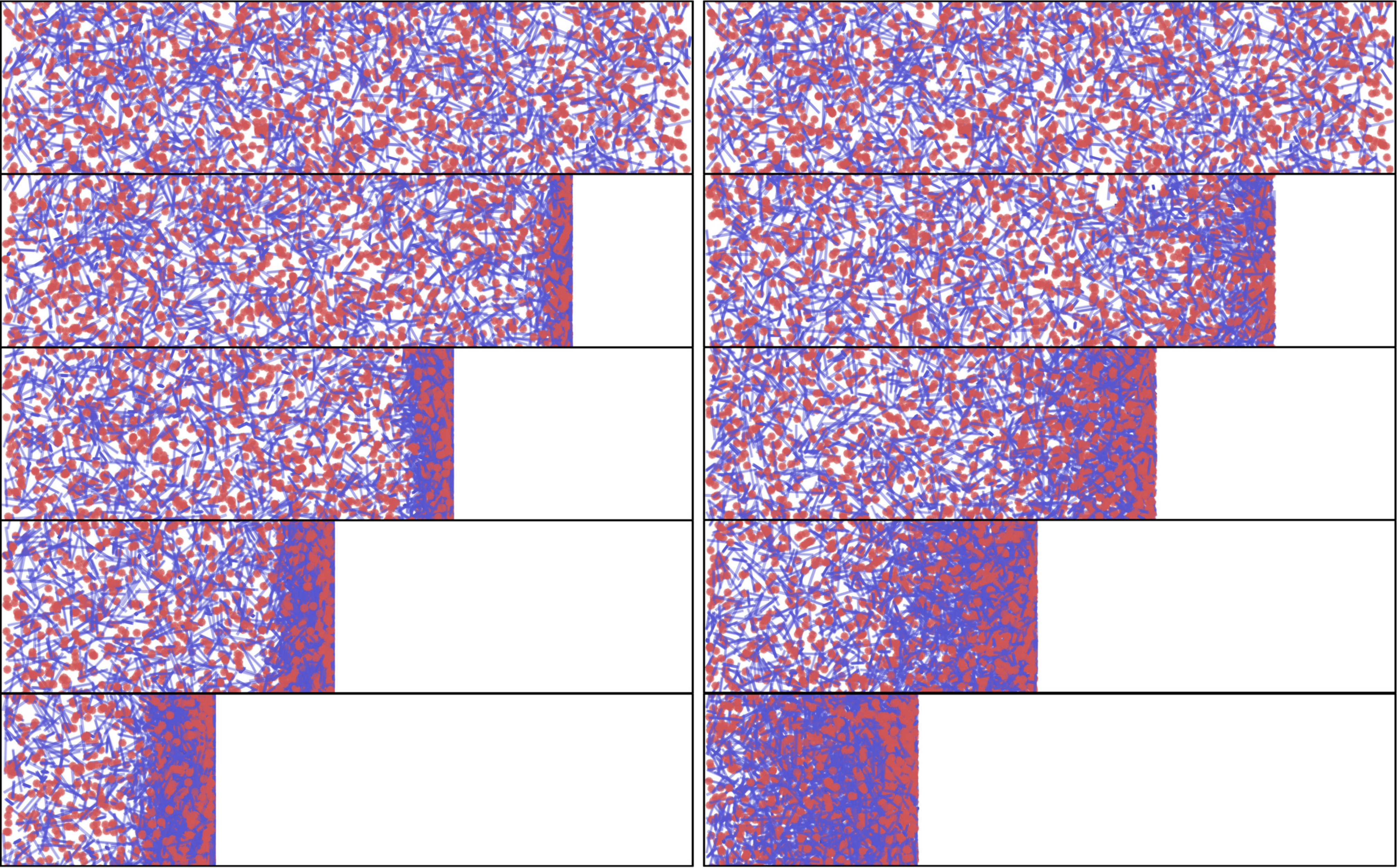}
\caption{Snapshots of the drying film for for system S5 (solid spheres + long rods).}
\label{fg:snapshots_sys_5}
\end{figure}

\begin{figure}[htb]
\includegraphics[width = 0.8\textwidth]{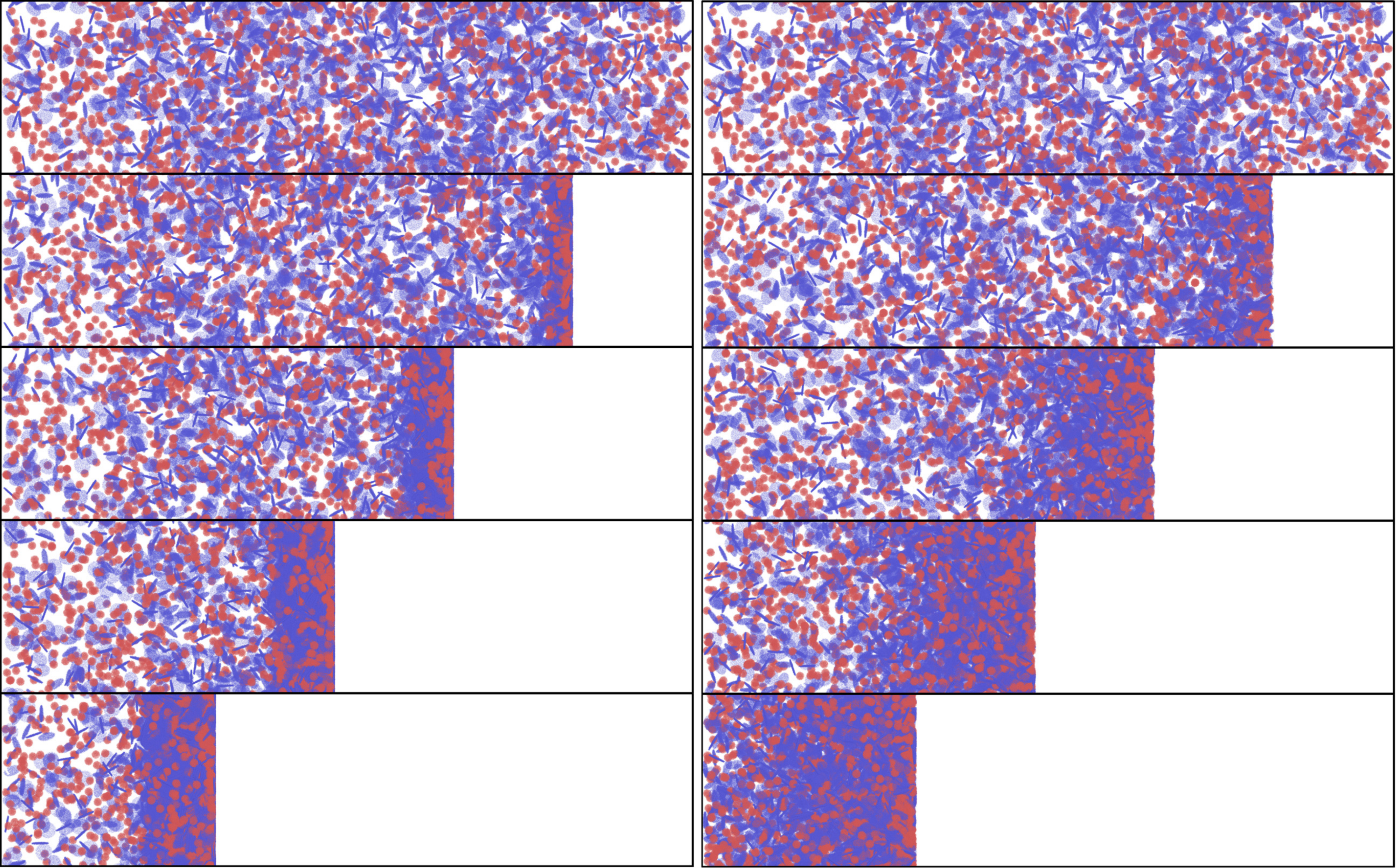}
\caption{Snapshots of drying films for system S6 (solid spheres + thin disks).}
\label{fg:snapshots_sys_6}
\end{figure}

\begin{figure}[htb]
\includegraphics[width = 0.8\textwidth]{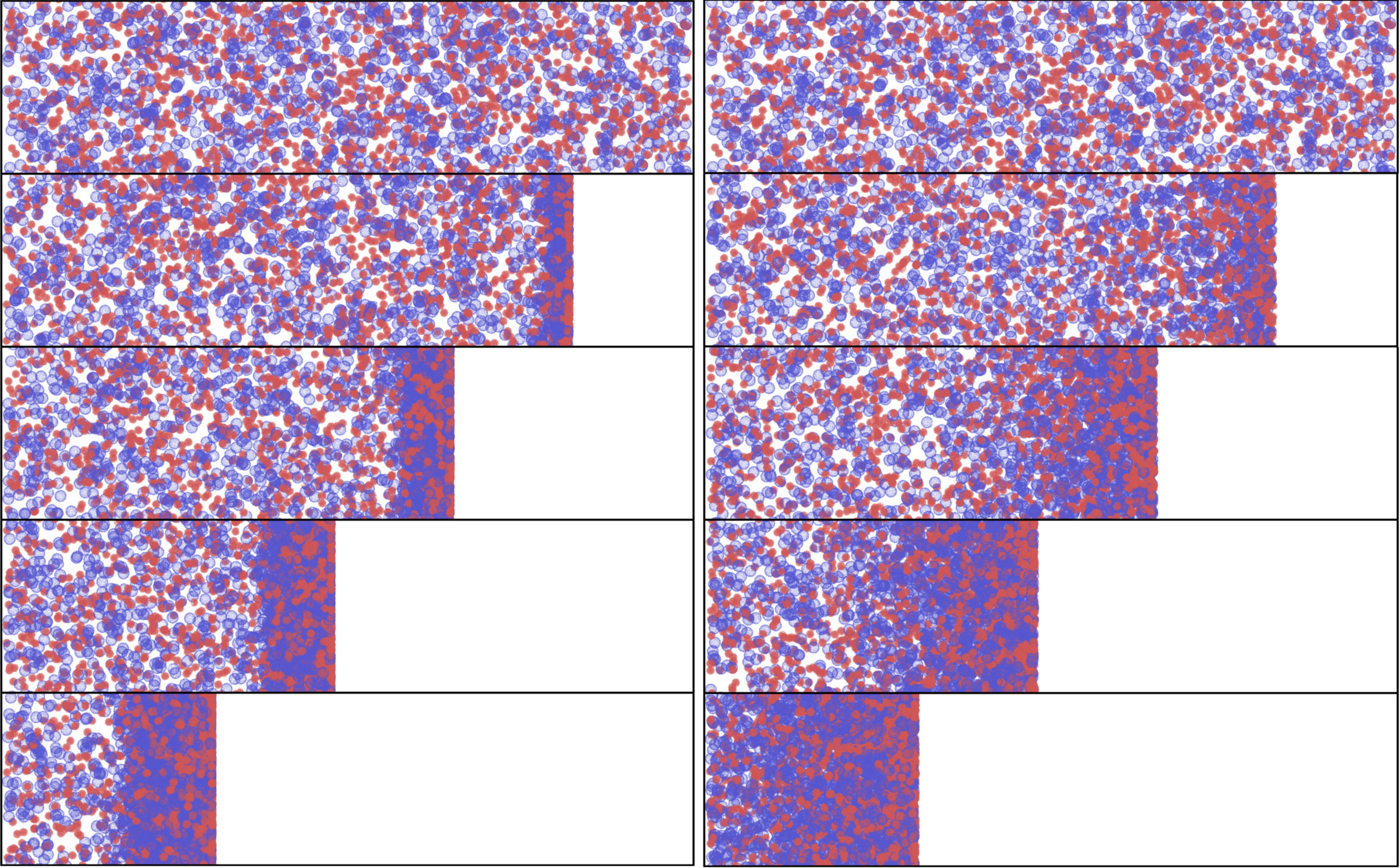}
\caption{Snapshots of the drying film for system S7 (solid spheres + hollow spheres). }
\label{fg:snapshots_sys_7}
\end{figure}

\begin{figure}[htb]
\includegraphics[width = 0.8\textwidth]{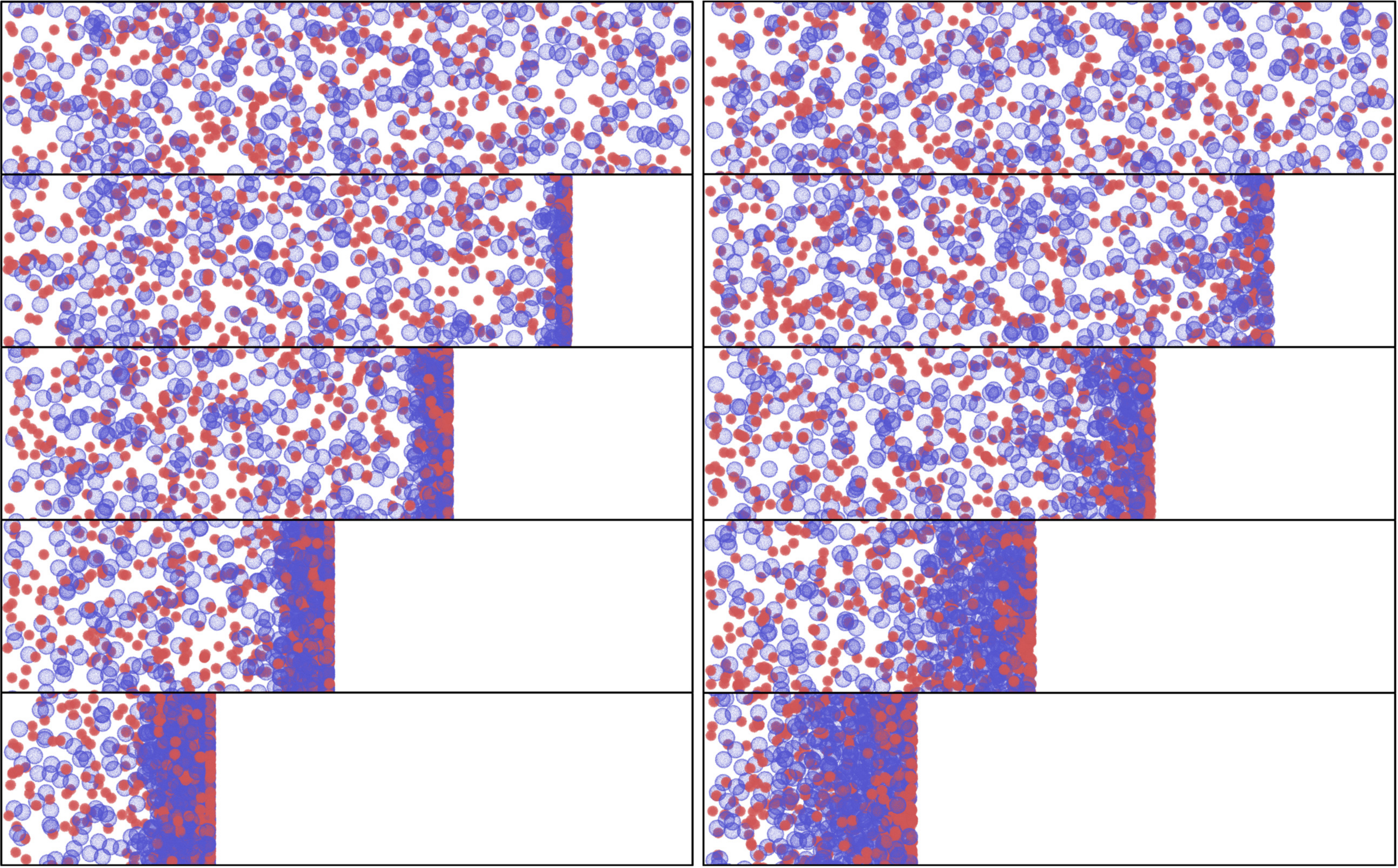}
\caption{Snapshots of the drying film for system S8 (larger solid spheres + larger hollow spheres).}
\label{fg:snapshots_sys_8}
\end{figure}

\begin{figure}[htb]
\includegraphics[width = 0.45\textwidth]{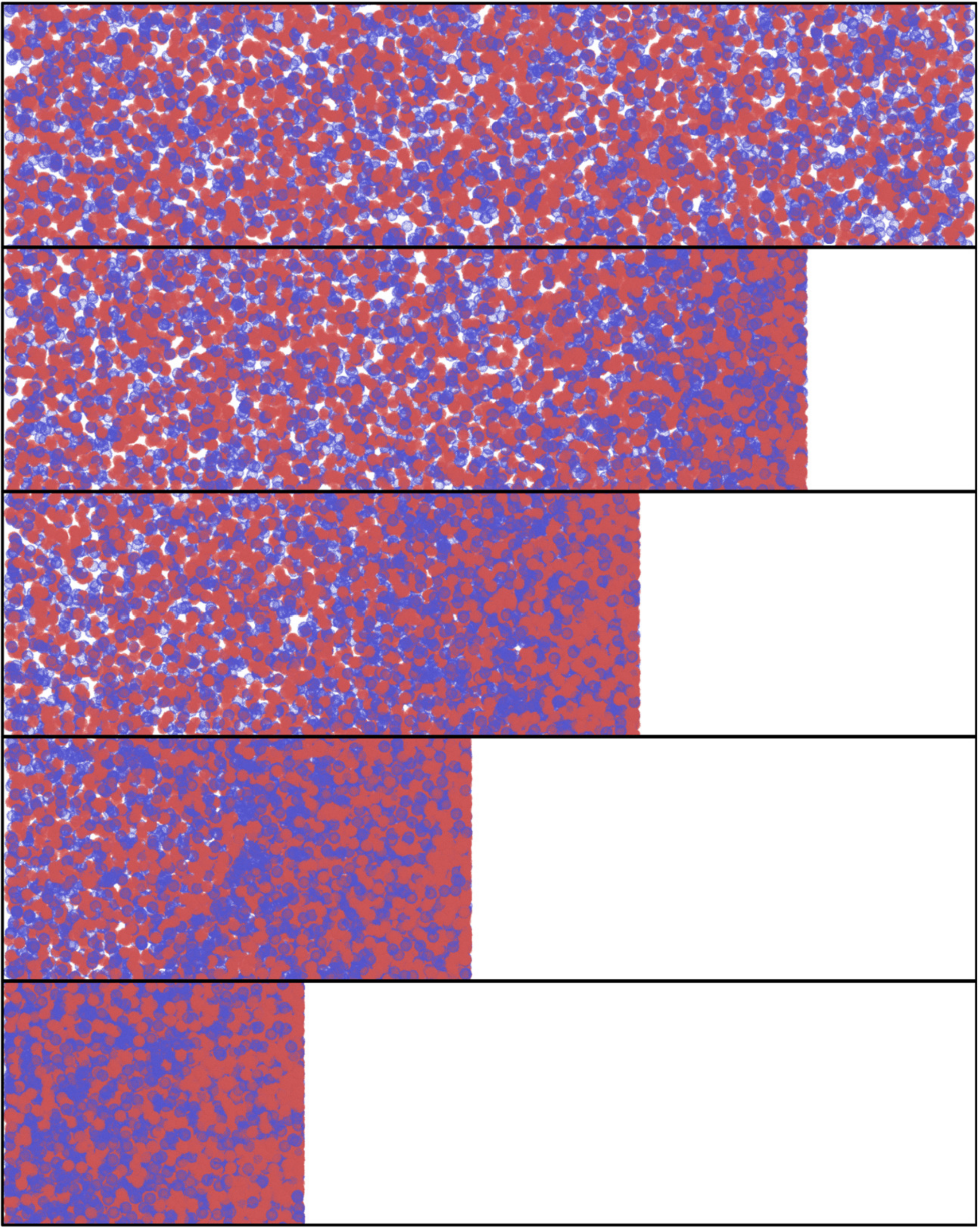}
\caption{Snapshots of the drying film for system S9 (solid spheres + hollow spheres with the same outer radius) under the {\bf slow-drying} condition with $v_{e} = 1.311\times 10^{-3}\sigma/\tau$.}
\label{fg:snapshots_sys_9_slow}
\end{figure}

\begin{figure}[htb]
\includegraphics[width = 0.45\textwidth]{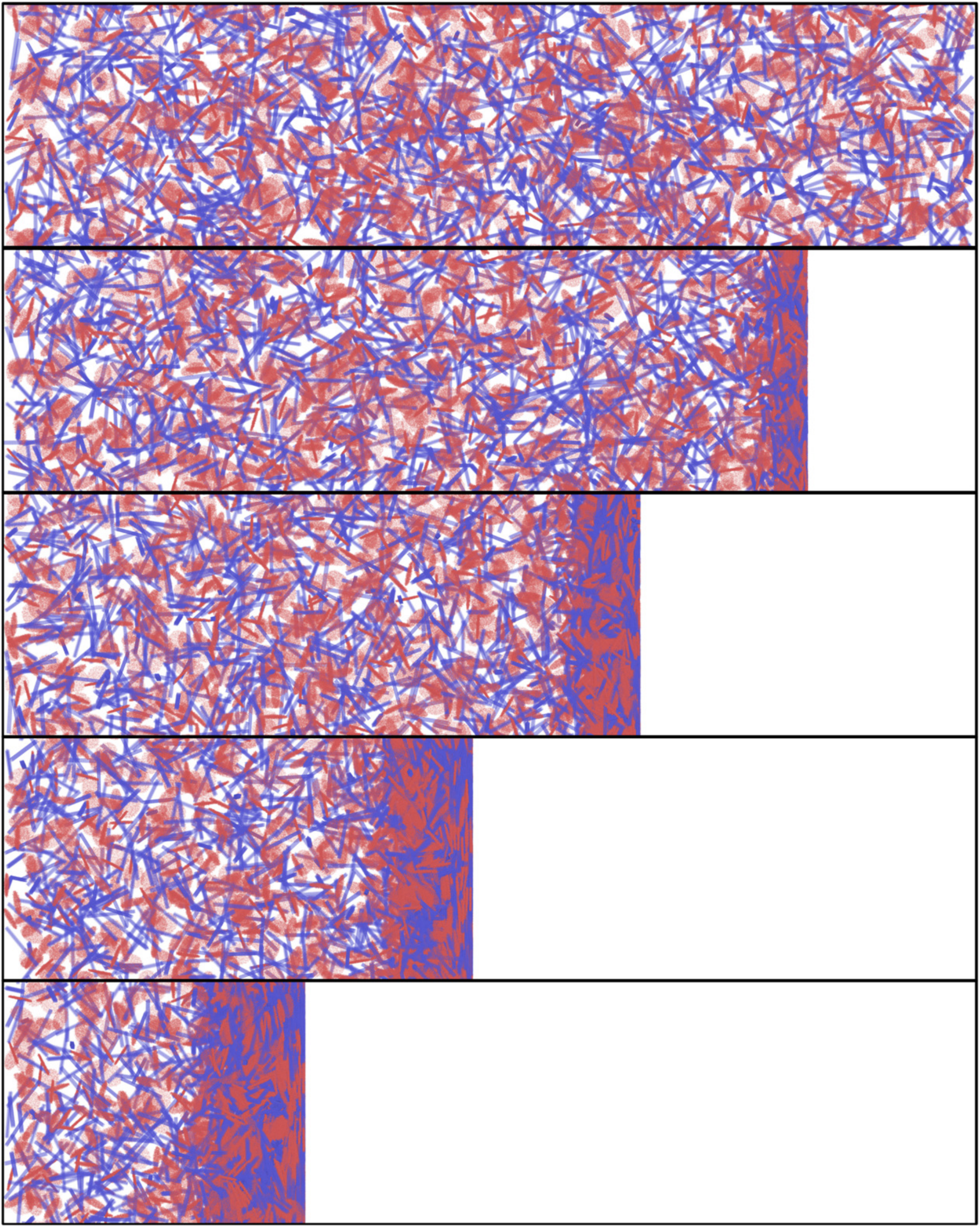}
\caption{Snapshots of the drying film for the mixture of thin
 disks (red) and long rods (blue) under the {\bf fast-drying} condition with $v_{e} = 1.311\times 10^{-2}\sigma/\tau$.}
\label{fg:snapshots_sys_10_fast}
\end{figure}

\FloatBarrier 

\section{Snapshots of Particle Distributions under the Slow-Drying Condition}

\begin{figure*}[htb]
  \includegraphics[width=1.0\textwidth]{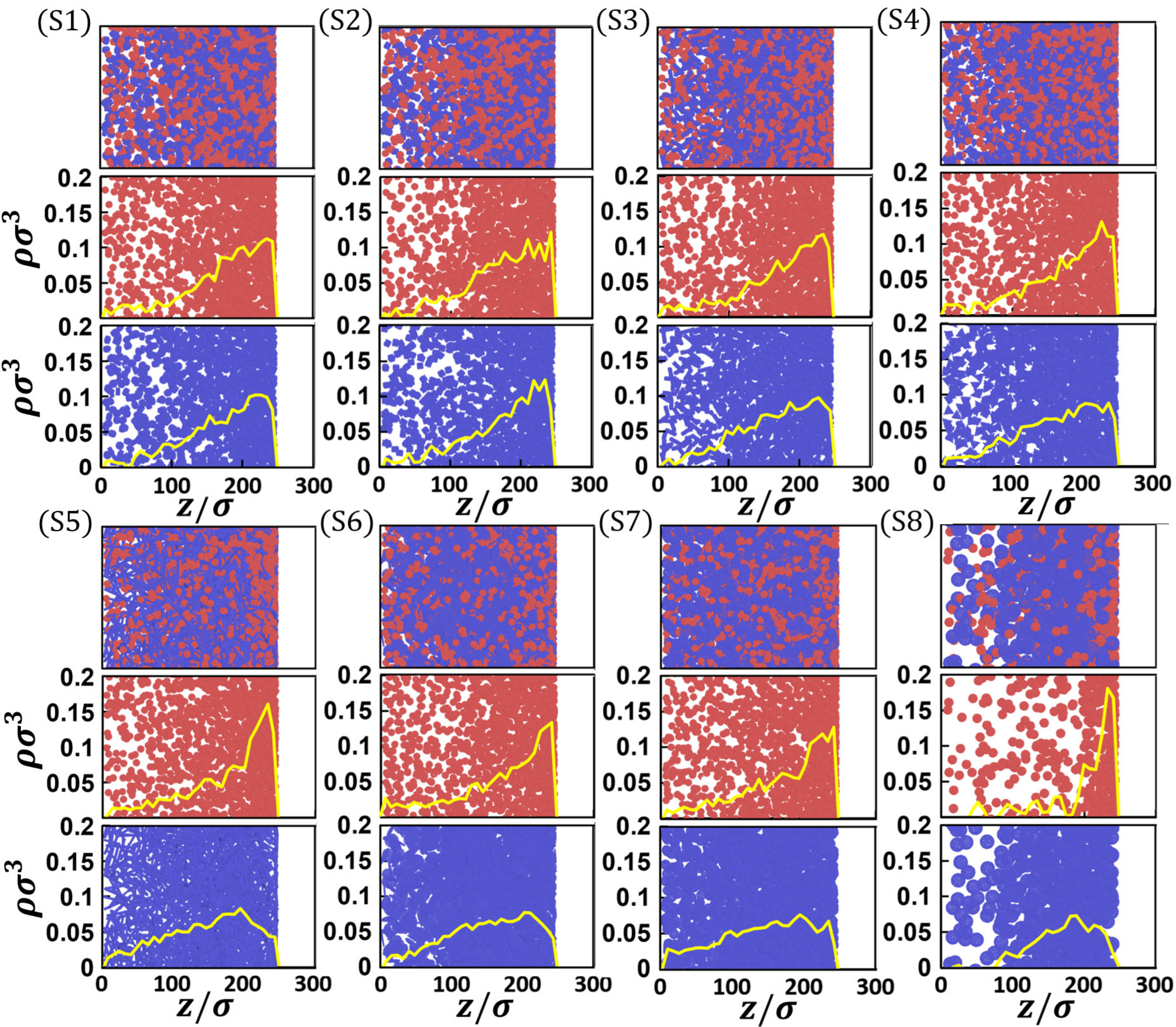}
  \caption{Projections, into the $xz$ plane, of particle distribution in the drying film of a thickness of $\sim 240\sigma$ for all systems discussed in the main text under the \textbf{slow-drying} condition with $v_{e} = 1.311\times 10^{-3}\sigma/\tau$. Each subset contains three images: mixture (top), solid spheres (middle), and aspherical particles (bottom). The yellow lines indicate the number density of the corresponding type of particles in the film along the direction of drying.}
  \label{fg:snapshot_film_240sigma_slow}
\end{figure*}

In this section. We include the end-stage snapshots in systems S1-S8 (see Table 1 of the main text) for the mixture (top subfigure), as well as for each type of particles in the mixture (middle and bottom subfigures) and their density profiles (yellow lines) along the direction of drying. Fig.~\ref{fg:snapshot_film_240sigma_slow} thus corresponds to Fig.~2 of the main text but under the slow-drying condition with $v_{e} = 1.311\times 10^{-3}\sigma/\tau$. The results are consistent with Fig.~2 of the main text, which is for the fast-drying condition; both follow the order ranked by the sphericity of the aspherical particles in the mixture. For systems S5 and S6, the density of the aspherical particles peaks at a location away from the receding interface. Therefore, they can be classified as stratified systems. However, system S4, which is marginally stratified under the fast-drying rate, does not show evidence of stratification under the slow-drying condition, as shown in Fig.~\ref{fg:snapshot_film_240sigma_slow}. Systems S7 and S8, in which the particles have comparable diffusion coefficients, show strong stratification due to the large surface area contrast between the particles.   

\FloatBarrier 

\section{Diffusion Coefficients of Composite Particles in Mixtures}

In this section, we include results on the mean-square displacement, $\langle (\Delta r)^2\rangle$, calculated with independent molecular dynamics simulations and averaged over particles of the same type in each mixture during a period of $t$, which are used to calculate the diffusion coefficients of the particles via
\begin{equation}
    D = \lim_{t\rightarrow \infty} \frac{\langle (\Delta r)^2\rangle}{6t}~.
\end{equation}
The results on $D$ are reported in Table 1 of the main text. In these simulations, each mixture is placed in a cubic box with a side length of $200\sigma$. Periodic boundary conditions are applied in all directions. Other simulation conditions (volume fractions, temperature, and thermostat) are identical to those used for the equilibrium mixtures in Table 1 of the main text before drying.

\begin{figure}[htb]
\hspace{-1.2cm}
\includegraphics[width = 0.9\textwidth]{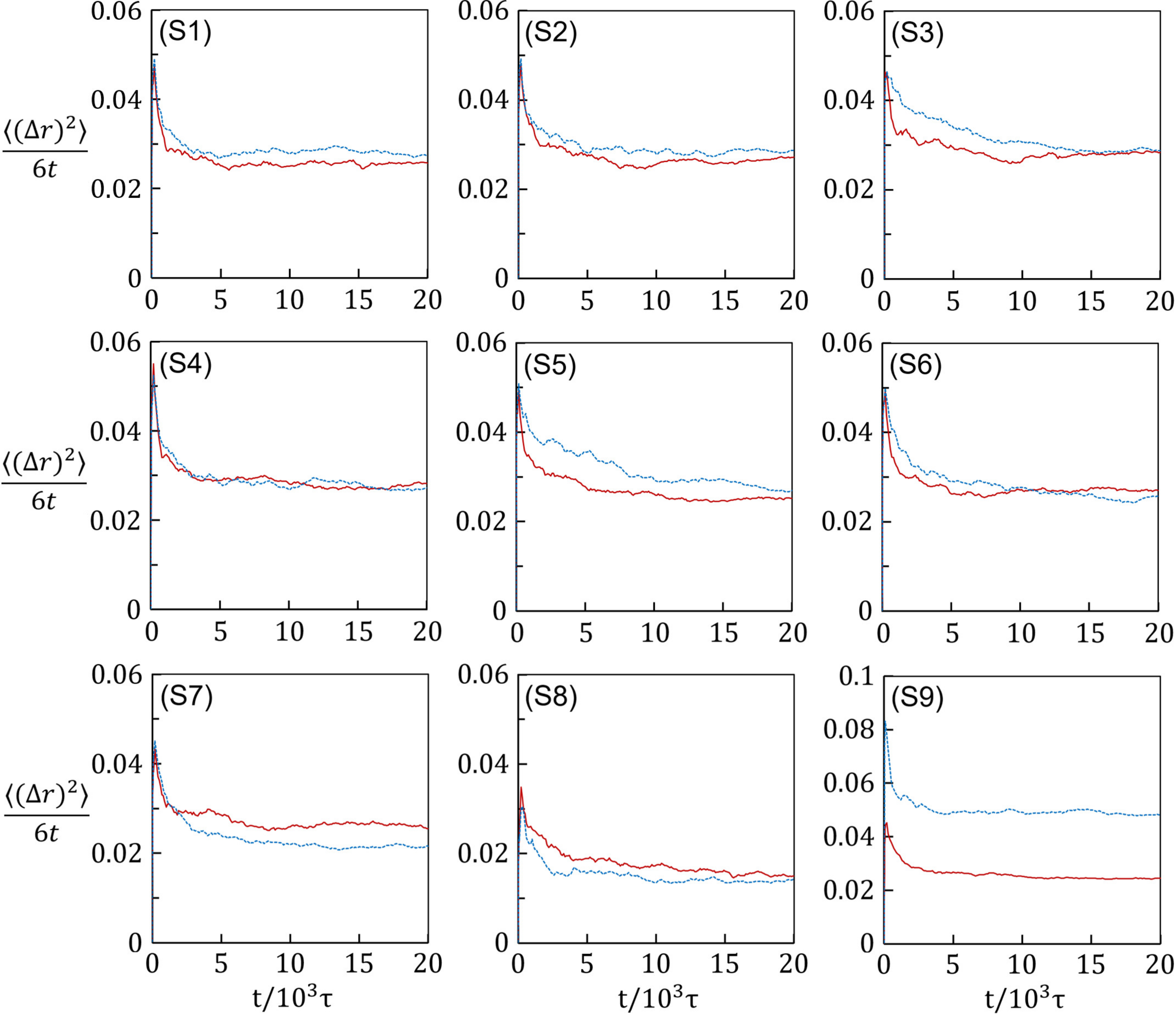}
\caption{Mean-square displacement, $\langle (\Delta r)^2\rangle$, averaged over particles of the same type in each mixture during a period of $t$ is divided by $6t$ and plotted against $t$ for the solid spheres (red lines) and the aspherical/hollow particles (dark blue lines).}
\label{fg:diffusion}
\end{figure}

\end{document}